\documentclass{LMCS}

\usepackage{enumerate}
\usepackage{hyperref}

\usepackage{amssymb}
\usepackage{graphicx}

\usepackage{url}
\usepackage{times}
\usepackage{proof}
\usepackage{amsmath}

\usepackage{stmaryrd} 
\usepackage{amssymb}
\usepackage{amsbsy}
\usepackage{amstext}

\newcommand{\sem}[1]{\llbracket #1 \rrbracket}

\newcommand{\Dashv}{\mathrel{\reflectbox{$\vDash$}}}

\newcommand{\hide}[1]{}

\newcommand{\bla}{\mathit{bla}}

\newcommand{\points}{\!\mapsto\!}

\newcommand{\fin}{\mathit{fin}}

\newcommand{\blackslug}{\rule{7pt}{7pt}}

\def\eqalign#1{\null\,\vcenter{\openup\jot\mathsurround=0 pt
  \ialign{\strut\hfil$\displaystyle{##}$&$\displaystyle{{}##}$\hfil
      \crcr#1\crcr}}\,}

\theoremstyle{plain}\newtheorem{satz}[thm]{Satz}
\theoremstyle{plain}\newtheorem{lemma}[thm]{Lemma}
\theoremstyle{plain}\newtheorem{proposition}[thm]{Proposition}
\theoremstyle{plain}
\def\eg{{\em e.g.}}
\def\cf{{\em cf.}}

\def\doi{5 (2:4) 2009}
\lmcsheading%
{\doi}
{1--27}
{}
{}
{Sep.~\phantom{0}1, 2008}
{Apr.~23, 2009}
{}   

\begin{document}

\title[Footprints in Local Reasoning]{Footprints in Local Reasoning\rsuper*}

\author[M.~Raza]{Mohammad Raza\rsuper a}	
\address{{\lsuper{a,b}}Department of Computing, \\Imperial College London, 
180 Queen's Gate, London SW7 2AZ, UK\\}	
\email{\{mraza,pg\}@doc.ic.ac.uk}  

\author[P.~Gardner]{Philippa Gardner\rsuper b}	



\titlecomment{{\lsuper*}A Preliminary version of this paper appeared in the FOSSACS 2008 conference}


\begin{abstract}
  \noindent Local reasoning about programs exploits the natural local behaviour common in programs by focussing on the footprint - that part of the resource accessed by the program. We address the problem of formally characterising and analysing the notion of footprint for abstract local functions introduced by Calcagno, O'Hearn and Yang. With our definition, we prove that the footprints are the only essential elements required for a complete specification of a local function. We formalise the notion of small specifications in local reasoning and show that, for well-founded resource models, a smallest specification always exists that only includes the footprints. We also present results for the non-well-founded case. Finally, we use this theory of footprints to investigate the conditions under which the footprints correspond to the smallest safe states. We present a new model of RAM in which, unlike the standard model, the footprints of every program correspond to the smallest safe states. We also identify a general condition on the primitive commands of a programming language which guarantees this property for arbitrary models. 
\end{abstract}

\maketitle

\section{Introduction}

Local reasoning about programs focusses on the collection of resources
directly acted upon by the program. It has recently been introduced
and used to substantial effect in {\em local} Hoare reasoning about
memory update. Researchers previously used Hoare reasoning based on
First-order Logic to specify how programs interacted with the {\em
  whole} memory.  O'Hearn, Reynolds and Yang instead introduced {
  local} Hoare reasoning based on Separation Logic ~\cite{ORY01,IO01}.  The idea is to reason only about the local parts of the
memory---the {\em footprints}---that are accessed by a
program.  Intuitively, the footprints form the pre-conditions of the
{\em small} axioms, which provide the smallest complete specification
of the program. All the true Hoare triples are derivable from the
small axioms and the general  Hoare rules. In particular, the {\em
  frame rule}  extends the reasoning to properties about the
rest of the heap which has not been changed by the command.

O'Hearn, Reynolds and Yang originally introduced Separation Logic to
solve the problem of how to reason about the mutation of data
structures in memory.  They have applied their reasoning to several
memory models, including heaps based on pointer
arithmetic~\cite{ORY01}, heaps with permissions~\cite{BCOP05}, and the
combination of heaps with variable stacks which views variables as
resource~\cite{BCY05,PBC06}. In each case, the basic soundness and
completeness results for local Hoare reasoning are essentially the
same. For this reason, Calcagno, O'Hearn and Yang~\cite{COY07}
recently introduced abstract local functions over abstract resource
models which they call separation algebras. They generalised their specific examples of
local imperative commands and memory models in this abstract framework. They introduced
Abstract Separation Logic to provide local Hoare reasoning about such functions, and give general soundness and completeness results.

We believe that the general concept of a local function is a fundamental step towards
establishing the theoretical foundations of local reasoning, and
Abstract Separation Logic is an important generalisation of the local
Hoare reasoning systems now widely studied in the literature.
However,  Calcagno, O'Hearn and Yang  do not
characterise the  footprints and small axioms in this
general theory, which is a significant  omission.  O'Hearn,
Reynolds and Yang, in one of their first papers on the
subject~\cite{ORY01}, state the local reasoning viewpoint as:
\begin{quote}
 `to understand how a program works, it should be possible for
  reasoning and specification to be confined to the cells that the
  program actually accesses. The value of any other cell will
  automatically remain unchanged.'
\end{quote}
A complete understanding of the foundations of local Hoare reasoning
therefore requires a formal characterisation of the footprint notion. 
O'Hearn tried to formalise footprints in his work on Separation Logic (personal communication with O'Hearn).
His intuition was that the footprints should be the smallest states
on which the program is safe - the {\em safety footprint}, and that the {\em small axioms} arising from
these footprints should give rise to a complete specification using
the general rules for local Hoare reasoning.  However, Yang discovered that
this notion of footprint does not work, since it does not always
yield a {\em complete} specification for the  program. Consider the program\footnote{Yang's example was the
  `allocate-deallocate-test' program \emph{ADT} ::= `x := new();dispose(x); 
 if \;(x=1) \; then \; z:=0 \; else \; z:=1;x=0'. Our  \emph{AD} program
 provides  a
 more standard example of program behaviour.}
\[
\mathit{AD} ::= \quad x := new();dispose(x)\]
This \emph{allocate-deallocate} program allocates a new cell, stores its address value in the stack variable $x$, and
then deallocates the cell. It is local because all its atomic constituents are
local. This tiny  example  captures the
essence of  a
common type of  program; there are many programs
which, for example,    create  a
list, work on the list, and then destroy the list.

The
smallest heap on which the \emph{AD} program  is safe is the empty heap
$emp$. The  specification using  this pre-condition is: 
\begin{eqnarray}
\{emp\} \quad \mathit{AD} \quad \{emp\}
\end{eqnarray}
We can extend our reasoning to  larger heaps by applying  the frame
rule: for example, extending  to a one-cell
heap with arbitrary address $l$ and value $v$ gives
\begin{eqnarray}
\{l \mapsto v \} \quad \mathit{AD} \quad \{l \mapsto v\}
\end{eqnarray}
However, axiom (1) does not give the complete
specification of the \emph{AD} program. In fact, it captures very little of
 the spirit of allocation followed by de-allocation.
For example, the following triple is also true:
\begin{eqnarray}
\{l \mapsto  v\} \quad \mathit{AD} \quad \{l \rightarrow v \wedge x \neq l\} 
\end{eqnarray}
This triple (3)   is true because, if $l$ is already 
allocated, then the new address cannot be $l$ and hence $x$ cannot be
$l$. It cannot be derived from (1). However, the combination of axiom
(1) and axiom (3) for arbitrary one-cell heaps does provide the
smallest complete specification. 
This example illustrates that O'Hearn's intuitive
view of  the footprints as the minimal safe states just does not work
 for common  imperative programs.

 In this paper, we introduce the formal definition of the footprint of a
 local function that does yield a complete specification for the function. For our \emph{AD} example, our definition identifies $emp$
 and the arbitrary one-cell heaps $l \mapsto v$ as footprints, as expected. We prove the general result that, for any local function, 
 the footprints  are the
 only elements which are {\em essential} to specify completely the
 behaviour of this function.  
 
 We then investigate the question of {\em sufficiency}. For well-founded resource, we show that the footprints are also always sufficient: that is, a complete
 specification always exists that only uses the footprints. We also explore
 results for the non-well-founded case, which depend on the presence of {\em negativity}. A resource has negativity if it is possible to combine two non-unit elements to get the unit, which is like taking two non-empty pieces of resource and joining them to get nothing. For non-well-founded models without
 negativity, such as
 heaps with infinitely divisible fractional permissions, either the
 footprints are sufficient (such as for the \emph{write} command in the
 permissions model) or there is no smallest complete
 specification (such as for the {\em read} command in the permissions model). For models with
 negativity, such as the integers under addition, we show that there do exist smallest complete specifications based
 on elements that are not essential and hence not footprints.
 
In the final section, we apply our theory of footprints to the issue of regaining the safety footprints. We address a question that arose from discussions with O'Hearn and Yang, which is whether there is an alternative model of RAM in which the safety footprint does correspond to the actual footprint, yielding complete specifications. We present such a model based on an examination of the cause of the \emph{AD} problem in the original model. We prove that in this new model the footprint of {\em every} program, including \emph{AD}, does correspond to the safety footprint. Moreover, we identify a general condition on the primitive commands of a programming language which ensures that this property holds in arbitrary models. 

A preliminary version of this paper was presented at the FOSSACS 2008 conference. The final section reports on work that is new to this journal version. This paper also contains the proofs which were excluded from the conference paper.

\section{Background}
\label{sec:separationalgebras}

The discussion in this paper is based on the framework introduced in \cite{COY07}, where the approach of local reasoning about programs with separation logic was generalised to local reasoning about \emph{local} functions that act on an abstract model of resource. Our objective in this work is to investigate the notion of footprint in this abstract setting, and this section gives a description of the underlying framework. 

\subsection{Separation Algebras and Local Functions}

We begin by describing separation algebras, which provide a model of resource which generalises over the specific heap models used in separation logic works. Informally, a separation algebra models resource as a set of elements that can be `glued' together to create larger elements. The `glueing' operator satisfies properties in accordance with this resource intuition, such as commutativity and associativity, as well as the cancellation property which requires that, if we are given an element and a subelement, then `ungluing' that subelement gives us a unique element. 

\begin{defi}[Separation Algebra]
A {\bf separation algebra} is a cancellative, partial commutative
monoid $(\Sigma,\bullet,u)$, where $\Sigma$ is a set and $\bullet$
is a partial binary operator with unit $u$. The operator satisfies the familiar axioms of associativity,
commutativity and unit, using a partial equality on $\Sigma$ where either both sides are defined and equal, or both are
undefined. It also satisfies the cancellative property stating that,
for each $\sigma \in \Sigma$, the partial function
$\sigma\bullet (\cdot ): \Sigma \points \Sigma $ is  injective. 
\end{defi}

We shall sometimes overload notation, using $\Sigma$ to denote the separation algebra $(\Sigma,\bullet,u)$. Examples of separation algebras include
multisets with union and unit $\emptyset$, the natural numbers with addition and unit $0$, heaps as finite partial functions from locations
to values (~\cite{COY07} and example \ref{locsepexamples}), heaps with
permissions ~\cite{COY07,BCOP05}, and
the combination of heaps and variable stacks enabling us to model programs with variables as local functions (~\cite{COY07}, ~\cite{PBC06} and example \ref{locsepexamples}). These examples all have an intuition of resource, with $\sigma_1 \bullet \sigma_2$
intuitively giving more resource than just $\sigma_1$ and $ \sigma_2$
for $\sigma_1 , \sigma_2 \neq u$. However, notice that the general notion of a separation algebra also permits  examples which may not have this resource intuition, such as  $\{a,u\}$ with $a \bullet a = u$. Since our aim is to investigate general properties of local reasoning, our inclination is to impose minimal restrictions on what counts as resource and to work with a simple definition of a separation algebra.

\hide{ These
examples 
have an
underlying notion of resource, with $\sigma_1 \bullet \sigma_2$
intuitively giving 
 more resource than just $\sigma_1$ and $ \sigma_2$
for $\sigma_1 , \sigma_2 \neq u$. An example
which does not  fit this resource intuition is the  separation algebra
$\{a,u\}$ with $a \bullet a =
u$. 
}

\begin{defi}[Separateness and substate]
Given a separation algebra $(\Sigma,\bullet,u)$, the {\bf separateness} ($\#$) relation between two states $\sigma_0, \sigma_1 \in \Sigma$ is given by $\sigma_0\# \sigma_1 \;\mbox{iff}\;  \sigma_0 \bullet \sigma_1\;\mbox{is defined}$. The {\bf substate} ($\preceq$) relation is given by $\sigma_0\preceq \sigma_1 \;\mbox{iff}\;  \exists \sigma_2.\, \sigma_1=\sigma_0\bullet \sigma_2$. We write $\sigma_0\prec \sigma_1$ when $\sigma_0\preceq \sigma_1$ and $\sigma_0 \neq \sigma_1$.
\end{defi}

\begin{lemma}[Subtraction]
\label{subtraction}
For $\sigma_1, \sigma_2 \in \Sigma$, if $\sigma_1 \preceq \sigma_2$ then there exists a unique element denoted $\sigma_2 - \sigma_1 \in \Sigma$, such that
$(\sigma_2 - \sigma_1) \bullet \sigma_1 = \sigma_2$.
\end{lemma}
\proof
Existence follows by definition of $\preceq$. For uniqueness, assume there exist $\sigma', \sigma'' \in \Sigma$ such that $\sigma'  \bullet \sigma_1 = \sigma_2$ and $\sigma'' \bullet \sigma_1 = \sigma_2$. Then we have  $\sigma'  \bullet \sigma_1 = \sigma'' \bullet \sigma_1$, and thus by the cancellation property we have $\sigma'  = \sigma''$.
\qed

We consider functions on separation algebras that generalise imperative programs operating on heaps. Such programs can behave non-deterministically, and can also \emph{fault}. To model non-determinism, we consider functions from a separation algebra $\Sigma$ to its powerset $\mathcal{P}(\Sigma)$. To model faulting, we add a special top element $\top$ to the powerset. We therefore consider total functions of the form $f:\Sigma\rightarrow \mathcal{P}(\Sigma)^\top$. On any element of $\Sigma$, the function can either map to a set of elements, which models \emph{safe} execution with non-deterministic outcomes, or to $\top$, which models a faulting execution. Mapping to the empty set represents divergence (non-termination).

\begin{defi}
The standard subset relation on the powerset is extended to $\mathcal{P}(\Sigma)^\top$ by defining $p  \sqsubseteq \top$ for all $p \in \mathcal{P}(\Sigma)^\top$. The binary operator $\ast$ on $\mathcal{P}(\Sigma)^\top$ is given by
\begin{eqnarray*}
p*q &=& \{\sigma_0\bullet \sigma_1 \mid \sigma_0 \# \sigma_1 \wedge
\sigma_0 \in p \wedge \sigma_1 \in q\} \quad \mathit{if}\;  p,  q \in \mathcal{P}(\Sigma)\\
  &=& \top \quad otherwise
\end{eqnarray*}
$\mathcal{P}(\Sigma)^\top$ is a total commutative monoid under $\ast$ with unit $\{u\}$. 
\end{defi}

\begin{defi}[Function ordering]
For functions $f, g:\Sigma\rightarrow \mathcal{P}(\Sigma)^\top$, $f \sqsubseteq g$ iff $f(\sigma) \sqsubseteq g(\sigma)$ for all $\sigma \in \Sigma$.
\end{defi}

\hide{
A \emph{local} function $f$ on separation algebra $\Sigma$ is a function
whose behaviour on an element $\sigma $ is closely linked with its
behaviour on the subelements.  If its action is to fault on $\sigma$,
then it must fault on all its subelements. The intuition is that if
$\sigma$ does not have enough resource, then neither does its
subelements. However, if $f$s action on a subelement $\sigma'$ returns
a result, then its action on $\sigma$ cannot fault (safety monotonicity
property)  and the result is is bounded by the result on $\sigma'$
and the extra resource $\sigma - \sigma'$. 
}

We shall only consider functions that are \emph{well-behaved} in the sense that they act \emph{locally} with respect to resource. For imperative commands on the heap model, the locality conditions were first characterised in \cite{YO02}, where a soundness proof for local reasoning with separation logic was demonstrated for the specific heap model. The conditions identified were 
\begin{enumerate}[$\bullet$]
\item \emph{Safety monotonicity}: if the command is safe on some heap, then it is safe on any larger heap.
\item \emph{Frame property}: if the command is safe on some heap, then in any outcome of applying the command on a larger heap, the additional heap portion will remain unchanged by the command.
\end{enumerate}

In \cite{COY07}, these two properties were amalgamated and formulated for abstract functions on arbitrary separation algebras.

\begin{defi}[Local Function]
\label{def:localaction}
A {\bf  local function on $\Sigma$}
is a total function $f:\Sigma\rightarrow \mathcal{P}(\Sigma)^\top$
which satisfies  the {\bf  locality condition}\/: 
\[\sigma  \# \sigma'\;\;\mbox{implies}\;\; f(\sigma' \bullet \sigma) \sqsubseteq \{\sigma'\} * f(\sigma)\]
We let $LocFunc$ be the set of local functions on $\Sigma$.
\end{defi}

Intuitively, we think of a command to be local if, whenever the
command executes safely on any resource element, then the command will
not `touch' any additional resource that may be added. Safety
monotonicity follows from the above definition because, if $f$ is safe
on $\sigma$ ($f(\sigma) \sqsubset \top$), then it is safe on any
larger state, since $f(\sigma'\bullet\sigma) \sqsubseteq \{\sigma'\} *
f(\sigma) \sqsubset \top$.

The frame property follows by the fact that the additional state
$\sigma'$ is preserved in the output of
$f(\sigma'\bullet\sigma)$. Note, however, that the $\sqsubseteq$
ordering allows for reduced non-determinism on larger states. This,
for example, is the case for the $AD$ command from the introduction
which allocates a cell, assigns its address to stack variable $x$, and
then deallocates the cell. On the empty heap, its result would allow
all possible values for variable $x$. However, on the larger heap
where cell 1 is already allocated, its result would allow all values
for $x$ except 1, and we therefore have a more deterministic outcome
on this larger state.

\begin{lemma}
Locality is preserved under sequential composition, non-deterministic choice and Kleene-star, which are defined as 
\[\eqalign{
(f;g)(\sigma)
&= \left\{ \begin{array}{ll}
\begin{array}{l} \top \end{array} &  \begin{array}{l}\mbox{ if } f(\sigma) = \top  \end{array} \\
\begin{array}{l} \bigsqcup \{ g(\sigma') \mid \sigma' \in f(\sigma)\}  \end{array} & \begin{array}{l}\mbox{ otherwise} \end{array}
\end{array}\right.\cr
(f+g)(\sigma) &= f(\sigma) \sqcup g(\sigma)\cr
f^\ast(\sigma)&= \displaystyle\bigsqcup_{n} f^n(\sigma)\cr}
\]

\end{lemma}


\begin{exa}[Separation algebras and local functions]
\label{locsepexamples}
  \hspace{1cm}
\begin{enumerate}[(1)]
\item {\bf Plain heap model}. A simple example is the separation algebra of heaps $(H, \bullet, u_H)$, where $H = L \rightharpoonup_{\fin} Val$ are finite partial functions from a set of  locations $L$ to a set of values $Val$ with $L \subseteq Val$, the partial operator $\bullet$ is the union of partial functions with disjoint domains, and the unit $u_H$ is the function with the empty domain. For $h \in H$, let $dom(h)$ be the domain of $h$. We write $l \mapsto v$ for the partial function with domain $\{l\}$ that maps $l$ to $v$. For $h_1, h_2 \in H$, if $h_2 \preceq h_1$ then $h_1 - h_2 = h_1\!\mid_{dom(h_1) - dom(h_2)}$. An example of a local function is the $dispose[l]$ command that deletes the cell at location $l$:
\[
dispose[l](h) 
= \left\{ \begin{array}{ll} 
\begin{array}{l}\{h - (l \points v)\} \end{array} & \; h \succeq (l \points v) \\
\begin{array}{l} \top \end{array} & \; \mbox{otherwise} 
\end{array}\right.
\]
The function is local: if $h \not \succeq (l \points v)$ then $dispose[l](h) = \top$, and $dispose[l](h' \bullet h) \sqsubseteq \top$. Otherwise, $dispose[l](h' \bullet h) 
= \{(h'\bullet h) - (l \points v)\} 
\sqsubseteq  \{h'\} * \{h - (l \points v)\}  
= \{h'\} * dispose[l](h)$. 

\item {\bf Heap and stack}. There are two approaches to modelling the
  stack in the literature. One is to treat the stack as a total
  function from variables to values, and only combine two heap and
  stack pairs if the stacks are the same. The other approach, which we
  use here, is to allow splitting of the variable stack and treat it
  as part of the resource. We can incorporate the variable stack into
  the heap model by using the set $H = L \cup Var
  \rightharpoonup_{\fin} Val$, where $L$ and $Val$ are as before and
  $Var$ is the set of stack variables $\{x, y, z, ...\}$. The
  $\bullet$ operator combines heap and stack portions with disjoint
  domains, and is undefined otherwise. The unit $u_H$ is the function
  with the empty domain which represents the empty heap and empty
  stack. Although this approach is limited to disjoint reference to
  stack variables, this constraint can be lifted by enriching the
  separation algebra with \emph{permissions} \cite{BCOP05}. However,
  this added complexity using permissions can be avoided for the
  discussion in this paper. For a state $h \in H$, we let $loc(h)$ and
  $var(h)$ denote the set of heap locations and stack variables in the
  domain of $h$ respectively. In this model we can define the
  allocation and deallocation commands as
\[\quad\enspace\enspace\eqalign{
new[x](h)
&= \left\{ \begin{array}{ll}
\begin{array}{l}\{ h' \bullet x \points l \bullet l \points w \mid w \in Val, l \in L \backslash loc(h')\} \end{array} &  \begin{array}{l}h = h'\bullet x \points v \end{array} \\
\begin{array}{l}\top \end{array} & \begin{array}{l}\mbox{otherwise} \end{array}
\end{array}\right.\cr 
dispose[x](h)
&= \left\{ \begin{array}{ll}
\begin{array}{l}\{ h' \bullet x \points l \} \end{array} &  \begin{array}{l}h = h'\bullet x \points l \bullet l \points v  \end{array} \\
\begin{array}{l}\top \end{array} & \begin{array}{l}\mbox{otherwise} \end{array}
\end{array}\right.\cr} 
\]

\noindent Commands for heap mutation and lookup can be defined as 
\[\eqalign{ 
mutate[x,v](h)
&= \left\{ \begin{array}{ll}
\begin{array}{l}\{ h' \bullet x \points l \bullet l \points v \} \end{array} &  \begin{array}{l}h = h'\bullet x \points l\bullet l \points w \end{array} \\
\begin{array}{l}\top \end{array} & \begin{array}{l}\mbox{otherwise} \end{array}
\end{array}\right.\cr
lookup[x,y](h)
&= \left\{ \begin{array}{ll}
\begin{array}{l}\{ h' \bullet x \points l \bullet l \points v \bullet y \points v \} \end{array} &  \begin{array}{l}h = h'\bullet x \points l\bullet l \points v \bullet y \points w \end{array} \\
\begin{array}{l}\top \end{array} & \begin{array}{l}\mbox{otherwise} \end{array}
\end{array}\right.\cr}
\]
The \emph{AD} command described in the introduction, which is the composition $new[x];dispose[x]$, corresponds to the following local function
\[
AD(h)
= \left\{ \begin{array}{ll}
\begin{array}{l}\{h' \bullet x \points l \mid l \in L\backslash loc(h')\} \end{array} &  \begin{array}{l}h = h'\bullet x \points v \end{array} \\
\begin{array}{l}\top \end{array} & \begin{array}{l}\mbox{otherwise} \end{array}
\end{array}\right.\qquad\enspace
\]
Note that in all cases, any stack variables that the command refers to should be in the stack in order for the command to execute safely, otherwise the command will be acting non-locally.

\item {\bf Integers}. The integers form a separation algebra under addition with identity 0. In this case we have that any `adding' function $f(x) = \{x + c\}$ that adds a constant $c$ is local, while a function that multiplies by a constant $c$, $f(x) = \{cx\}$, is non-local in general. However, the integers under multiplication also form a separation algebra with identity 1, and in this case every multiplying function is local but not every adding function. This illustrates the point that the notion of locality of commands depends on the notion of separation of resource that is being used.  
\end{enumerate}
\end{exa}

\subsection{Predicates, Specifications and Local Hoare Reasoning}

We now present the local reasoning framework for local functions on separation algebras. This is an adaptation of Abstract Separation Logic \cite{COY07}, with some minor changes in formulation for the purposes of this paper. Predicates over separation algebras are treated simply as subsets of the separation algebra.  

\begin{defi}
\label{definition:predicates}
A {\bf predicate} $p$ over $\Sigma$ is an element of the powerset $\mathcal{P}(\Sigma)$.  
\end{defi}
Note that the top element $\top$ is not a predicate and that the $*$ operator, although defined on $\mathcal{P}(\Sigma)^\top \times \mathcal{P}(\Sigma)^\top \rightarrow \mathcal{P}(\Sigma)^\top$, acts as a binary connective on predicates. We have the distributive law for union that, for any  $X \subseteq \mathcal{P}(\Sigma)$,
$$(\bigsqcup X) * p = \bigsqcup \{ x*p\mid x\in X\} $$
The same is not true for intersection in general, but does hold for $precise$ predicates. A predicate is precise if, for any state, there is at most a single substate that satisfies the predicate. 
\begin{defi}[Precise predicate]
\label{precise}
A predicate $p \in \mathcal{P}(\Sigma)$ is {\bf precise} iff, for every $\sigma \in \Sigma$, there exists at most one $\sigma_p\in p$ such that $\sigma_p\preceq\sigma$.
\end{defi}

Thus, with precise predicates, there is at most a unique way to break a state to get a substate that satisfies the predicate. Any singleton predicate $\{\sigma\}$ is precise. Another example of a precise predicate is $\{l \points v \mid v \in Val\}$ for some $l$, while $\{l \points v \mid l \in L\}$ for some $v$ is not precise.    
\begin{lemma}[Precision characterization]
\label{precisecharacterization}
A predicate $p$ is precise iff, for all $X \subseteq \mathcal{P}(\Sigma)$, $(\bigsqcap X) * p = \bigsqcap \{ x * p \mid x\in X\}$
\end{lemma}

\proof
We first show the left to right direction. Assume $p$ is precise. We have to show that for all $X \subseteq \mathcal{P}(\Sigma)$, $(\bigsqcap X) * p = \bigsqcap \{ x * p \mid x\in X\}$. Assume $\sigma \in (\bigsqcap X) * p$. Then there exist $\sigma_1, \sigma_2$ such that $\sigma = \sigma_1\bullet\sigma_2$ and $\sigma_1 \in \bigsqcap X$ and $\sigma_2 \in p$. Thus for all $x \in X$, $\sigma \in x * p$, and hence $\sigma \in  \bigsqcap \{ x * p \mid x\in X\}$. Now assume $\sigma \in  \bigsqcap \{ x * p \mid x\in X\}$. Then $\sigma \in x * p$ for all $x \in X$. Hence there exists $\sigma_1 \preceq \sigma$ such that $\sigma_1 \in p$. Since $p$ is precise, $\sigma_1$ is unique. Let $\sigma_2 = \sigma - \sigma_1$. Thus we have $\sigma_2 \in x$ for all $x \in X$, and so $\sigma_2 \in \bigsqcap X$. Hence we have $\sigma \in (\bigsqcap X) * p$.

For the other direction, we assume that $p$ is not precise and show that there exists an $X$ such that $(\bigsqcap X) * p \neq \bigsqcap \{ x * p \mid x\in X\}$. Since $p$ is not precise, there exists $\sigma \in \Sigma$ such that, for two distinct $\sigma_1, \sigma_2 \in p$, we have $\sigma_1 \preceq \sigma$ and $\sigma_2 \preceq \sigma$. Let $\sigma'_1 = \sigma - \sigma_1$ and $\sigma'_2 = \sigma - \sigma_2$. Now let $X = \{\{\sigma'_1\},\{\sigma'_2\}\}$. Since $\sigma \in \{\sigma'_1\} * p$ and  $\sigma \in \{\sigma'_2\} * p$, we have  $\sigma \in  \bigsqcap \{ x * p \mid x\in X\}$. However, because of the cancellation property, we also have that $\sigma'_1 \neq \sigma'_2$, and so $(\bigsqcap X) * p = \emptyset * p = \emptyset$. Hence, $\sigma \not \in (\bigsqcap X) * p$, and we therefore have $(\bigsqcap X) * p \neq \bigsqcap \{ x * p \mid x\in X\}$.
\qed

\hide{
In the introduction we discussed the intuition of how the footprints are expected to correspond to the \emph{smallest} safe states. We will return to this point in section \ref{onestep}, employing the notion of a \emph{saturated} predicate, which is one that is always falsified on bigger states. Saturated can also be called `anti-intuitionistic', because an intuitionistic predicate is one that is never falsified on bigger states. Every precise predicate is also saturated. 

\begin{defi}[Saturated predicate]
\label{saturated}
A predicate $p \in \mathcal{P}(\Sigma)$ is {\bf saturated} if for every $\sigma \in p$, there is no $\sigma'$ in p such that $\sigma \prec \sigma'$.  
\end{defi}
}

Our Hoare reasoning framework is formulated with tuples of pre- and post- conditions, rather than the usual Hoare triples that include the function as in \cite{COY07}. In our case the standard triple shall be expressed as a function $f$ {\em satisfying} a tuple $(p, q)$, written $f \models (p, q)$. The reason for this is that we shall be examining the properties that a pre- and post- condition tuple may have with respect to a given function, such as whether a given tuple is complete for a given function. This approach is very similar to the notion of the \emph{specification statement} (a Hoare triple with a `hole') introduced in \cite{M88}, which is used in refinement calculi, and was also used to prove completeness of a local reasoning system in \cite{YO02}. 
\begin{defi}[Specification]
Let $\Sigma$ be a separation algebra. A {\bf statement} on $\Sigma$ is a tuple $(p, q)$, where $p, q \in \mathcal{P}(\Sigma)$ are predicates. A {\bf specification} $\phi$ on $\Sigma$ is a set of statements. We let $\Phi_{\Sigma} = \mathcal{P}(\mathcal{P}(\Sigma) \times \mathcal{P}(\Sigma))$ denote the set of all specifications on $\Sigma$. We shall exclude the subscript when it is clear from the context. The {\bf domain} of a specification is defined as $D(\phi) = \bigsqcup \{p \mid (p, q) \in \phi\}$. {\bf Domain equivalence} is defined as $\phi \cong_{D} \psi  \mbox{ iff }  D(\phi) = D(\psi)$.
\end{defi}
Thus the domain is the union of the preconditions of all the statements in the specification. It is one possible measure of \emph {size}: how much of $\Sigma$ the specification is referring to. We also adapt the notion of precise predicates to specifications.  
\begin{defi}
\label{precisesaturatedspecification}
A specification is precise iff its domain is precise.
\end{defi}


\begin{defi}[Satisfaction]
A local function $f$ satisfies a statement $(p,q)$, written $f \models (p,q)$, iff, for all $\sigma \in p$, $f(\sigma) \sqsubseteq q$. It satisfies a specification $\phi \in \Phi$, written $f \models \phi$, iff $f \models (p, q)$ for all $(p, q) \in \phi$.  
\end{defi}
\begin{defi}[Semantic consequence]
\label{semanticconsequence}
Let $p, q, r, s \in \mathcal{P}(\Sigma)$ and $\phi, \psi \in \Phi$. Each judgement $(p, q) \models (r, s), \phi \models (p, q)$, $(p, q) \models \phi$, and $\phi \models \psi$ holds iff all local functions that satisfy the left hand side also satisfy the right hand side.
\end{defi}
\begin{proposition}[Order Characterization]
\label{ordercharacterization}
$f \sqsubseteq g$ iff, for all $p,q \in  \mathcal{P}(\Sigma)$, $g \models (p, q)$ implies $f \models (p, q)$.\qed
\end{proposition}

For every specification $\phi$, there is a `best' local function satisfying $\phi$ (lemma \ref{blalemma}), in the sense that all statements that the best local function satisfies are satisfied by any local function that satisfies $\phi$. For example, in the heap and stack separation algebra of example \ref{locsepexamples}.2, consider the specification 
$$\phi_{new} = \{(\{x \points v\},\{x \points l \bullet l \points w \mid l \in L, w \in Val\}) \mid v \in Val\}$$
There are many local functions that satisfy this specification. Trivially, the local function that always diverges satisfies it. Another example is the local function that assigns the value $w$ of the newly allocated cell to be 0, rather than any non-deterministically chosen value. However, the best local function for this specification is the $new[x]$ function described in example \ref{locsepexamples}.2, as it can be checked that for any local function $f$ satisfying $\phi_{new}$, we have $f \sqsubseteq new[x]$. The notion of the best local function shall be used when addressing questions about completeness of specifications. It is adapted from \cite{COY07}, except that we generalise to the best local function of a specification rather than a single pre- and post-condition pair. 
\begin{defi}[Best local function]
For a specification $\phi \in \Phi$, the best local function of $\phi$, written $\bla[\phi]$, is
the function of type $\Sigma \rightarrow \mathcal{P}(\Sigma)^\top$  defined by
\[
\bla[\phi](\sigma) = \bigsqcap \{\{\sigma'\} * q \mid \sigma = \sigma'\bullet\sigma'', \sigma''\in p, (p, q) \in \phi\}
\]
\end{defi}

As an example, it can be checked that the best local function $\bla[\phi_{new}]$ of the specification $\phi_{new}$ given above is indeed the function $new[x]$ described in example \ref{locsepexamples}.2. The following lemma presents the important properties which characterise the best local function.

\begin{lemma}
\label{blalemma}
Let $\phi \in \Phi$. The following hold:
\begin{enumerate}[$\bullet$]
\item $\bla[\phi]$ is local
\item $\bla[\phi] \models \phi$ 
\item if $f$ is local and $f \models \phi$ then $f \sqsubseteq \bla[\phi]$
\end{enumerate}
\end{lemma} 

\proof
To show that $\bla[\phi]$ is local, consider $\sigma_1,\sigma_2$ such that $\sigma_1\#\sigma_2$.
We then calculate
\[\begin{array}{rll}
\bla[\phi](\sigma_1\bullet \sigma_2)\enspace
= & \bigsqcap \{\{\sigma'\} * q \mid \sigma_1\bullet \sigma_2 =\sigma'\bullet \sigma'', \sigma''\in p, (p, q) \in \phi\} \\
\sqsubseteq & \bigsqcap \{\{\sigma_1\bullet \sigma'''\} * q \mid \sigma_2=\sigma'''\bullet \sigma'', \sigma''\in p, (p, q) \in \phi\} \\
= & \bigsqcap \{\{\sigma_1\} * \{\sigma'''\} * q \mid \sigma_2=\sigma'''\bullet \sigma'', \sigma''\in p, (p, q) \in \phi\} \\
= & \{\sigma_1\} *  \bigsqcap \{\{\sigma'''\} * q \mid \sigma_2=\sigma'''\bullet \sigma'', \sigma''\in p, (p, q) \in \phi\} \\
= & \{\sigma_1\} * \bla[\phi](\sigma_2) 
\end{array}\]
In the second-last step we used the property that $\{\sigma_1\}$ is precise (lemma \ref{precisecharacterization}).\\

To show that $\bla[\phi]$ satisfies $\phi$, consider 
$(p, q) \in
\phi$ and $\sigma\in p$.  Then $\bla[\phi](\sigma) \sqsubseteq \{u\} *
q = q$.

For the last point, suppose $f$ is local and $f \models \phi$.
Then, for any $\sigma$ such that $\sigma=\sigma_1\bullet \sigma_2$ and $\sigma_2\in p$ and $(p, q) \in \phi$,
\[\begin{array}{lcll}
f(\sigma) &= & f(\sigma_1\bullet \sigma_2) \\
&\sqsubseteq & \{\sigma_1\} * f(\sigma_2) \\
&\sqsubseteq & \{\sigma_1\} * q \\
\end{array}\]
Thus $f(\sigma) \sqsubseteq \bla[\phi](\sigma)$.\\
In the case that there do not exist $\sigma_1,\sigma_2$ such that $\sigma=\sigma_1\bullet \sigma_2$ and $\sigma_2 \in D(\phi)$, then
\[\begin{array}{lcll}
\bla[\phi](\sigma) &= & \bigsqcap \emptyset \\
&= & \top \\
\end{array}\]
So in this case also $f(\sigma) \sqsubseteq \bla[\phi](\sigma)$.
\qed

\begin{lemma}
\label{blaconsequence}
For $\phi \in \Phi$ and $p, q \in \mathcal{P}(\Sigma)$, $\bla[\phi] \models (p, q) \Leftrightarrow \phi \models (p, q)$.
\end{lemma} 
\proof
\[\begin{array}{cll}
& \bla[\phi] \models (p, q) \\
\hbox to50 pt{\hfill}
\Leftrightarrow 
& \mbox{for all local functions } f,\; 
  f \models \phi \Rightarrow f \models (p, q) 
& \mbox{ (by lemma \ref{blalemma})} \\
\hbox to50 pt{\hfill}
\Leftrightarrow 
& \phi \models (p, q) 
& \mbox{ (by definition \ref{semanticconsequence})}.
  \hbox to51 pt{\hfill\qEd}
\end{array}
\]

\begin{figure}[t]
\hrule
\begin{center}
$\,$
$$
\begin{array}{cccc}
\infer 
{
 (p*r, q*r)
}
{ (p, q) }
&\quad
\infer
{
  (p', q')
}
{  p' \sqsubseteq p & (p,q) & q \sqsubseteq q'}
\quad&\quad
\infer
{
  \left(\bigsqcup_{i\in I} p_i, \bigsqcup_{i\in I} q_i\right)
}
{   (p_i,q_i),\;\mbox{all}\;i\in I}
\quad&\quad
\infer
{
  \left(\bigsqcap_{i\in I} p_i, \bigsqcap_{i\in I} q_i\right)
}
{   (p_i, q_i),\;\mbox{all}\;i\in I, I \neq \emptyset}
\\[2ex]
Frame & Consequence & Union & Intersection 
\end{array}
$$
\end{center}
\hrule
\caption{Inference rules for local Hoare reasoning}
\label{figure:rules}
\end{figure}
The inference rules of the proof system are given in figure \ref{figure:rules}. Consequence, union and intersection are adaptations of standard rules of Hoare logic. The frame rule is what permits local reasoning, as it codifies the fact that, since all functions are local, any assertion about a separate part of resource will continue to hold for that part after the application of the function. We omit the standard rules for basic constructs such as sequential composition, non-deterministic choice, and Kleene-star which can be found in \cite{COY07}. 
\begin{defi}[Proof-theoretic consequence]
\label{proofconsequence}
For predicates $p, q, r, s$ and specifications $\phi, \psi$, each of the judgements $(p, q) \vdash (r, s), \phi \vdash (p, q)$, $(p, q) \vdash \phi$, and $\phi \vdash \psi$ holds iff the right-hand side is derivable from the left-hand side by the rules in figure \ref{figure:rules}.
\end{defi}

The proof system of figure \ref{figure:rules} is sound and complete with respect to the satisfaction relation.

\begin{thm}[Soundness and Completeness]
\label{completeness}
$\phi \vdash (p, q) \Leftrightarrow \phi \models  (p, q)$
\end{thm}

\proof
Soundness can be checked by checking each of the proof rules in figure \ref{figure:rules}. The frame rule
is sound by the locality condition, and the others are easy to check. 

For completeness, assume we are given $\phi \models (p, q)$.  By lemma \ref{blaconsequence}, we have $\bla[\phi] \models (p, q)$.
So for all $\sigma \in p$, $\bla[\phi](\sigma) \sqsubseteq q$, which implies 
$$ \displaystyle\bigsqcup_{\sigma \in p} \bla[\phi](\sigma) \sqsubseteq q \quad (*) $$
Now we have the following derivation:

{\small
\[
\infer
{
	(p, q)
}
{
	\infer
	{
		(\displaystyle\bigsqcup_{\sigma \in p} \{\sigma\}, \displaystyle\bigsqcup_{\sigma \in p} \bla[\phi](\sigma))  
	}
	{	
		\infer
  		{
  			(\{\sigma\}, \bla[\phi](\sigma)) \quad \mbox{\tiny{for all $\sigma \in p$}}
  		}
  		{
 			\infer
  			{
				\big( \displaystyle\bigsqcap_{\substack{\sigma' \preceq \sigma \\ \sigma' \in r \\ (r, s) \in \phi}}  \{\sigma - \sigma'\} * \{\sigma'\}, \displaystyle\bigsqcap_{\substack{\sigma' \preceq \sigma \\ \sigma' \in r \\ (r, s) \in \phi}}  \{\sigma - \sigma'\} * s \big) \quad \mbox{\tiny{for all $\sigma \in p$}}
  			}	
  			{
  				\infer
  				{	
						( \{\sigma - \sigma'\} * \{\sigma'\}, \{\sigma - \sigma'\} * s) \quad \mbox{\tiny{for all $\sigma' \in r, (r, s) \in \phi, \sigma' \preceq \sigma, \sigma \in p$}}
  				}
  				{
  					\infer
  					{
						(\{\sigma'\}, s) \quad \mbox{\tiny{for all $\sigma' \in r, (r, s) \in \phi$}}
  					}
  					{
		  				\infer
		  				{
		  					(r, s) \quad \mbox{\tiny{for all $(r, s) \in \phi$}}
						}
						{
							\phi
						}  						
		  			}
  				}
  			}
  		}
	}
}
\]}
The last step in the proof is by $(*)$ and the rule of consequence. Note that the intersection rule can be safely applied because the argument of the intersection is necessarily non-empty (if it were empty then $\bla[\phi](\sigma) = \top$, which contradicts $\bla[\phi](\sigma) \sqsubseteq q$).
\qed

\section{Properties of Specifications}
\label{completespecifications}

We discuss certain properties of specifications as a prerequisite for our main discussion on footprints in Section~4. We introduce the notion of a {\em complete} specification for a local function, which is a specification from which follows every property that holds for the function. However, a function may have many complete specifications, so we introduce a canonical form for specifications. We show that of all the complete specifications of a local function, there exists a unique canonical complete specification for every domain. As discussed in the introduction, an important notion of local reasoning is the \emph{small specification} which completely describes the behaviour of a local function by mentioning only the footprint. Thus, as a prerequisite to investigating their existence, we formalise small specifications as complete specifications with the smallest possible domain. Similarly, we define {\em big} specifications as complete specifications with the biggest domain.

\begin{defi}[Complete Specification]
A specification $\phi \in \Phi$ is a {\bf complete specification} for $f$, written $complete(\phi, f)$, iff, for all $p, q \in \mathcal{P}(\Sigma)$,$ f \models (p, q) \Leftrightarrow \phi \models (p, q) $. Let {\bf $\Phi_{comp(f)}$} be the set of all complete specifications of f. 
\end{defi}
$\phi$ is complete for $f$ whenever the tuples that hold for $f$ are
{\em exactly} the tuples that 
 follow from $\phi$. This also means that any two complete specfications $\phi$ and $\psi$ for a local function are semantically equivalent, that is, $\phi \Dashv \vDash \psi$. The following proposition illustrates how the notions of best local action and complete specification are closely related. 
\begin{proposition}
\label{blaiffcomplete}
For all $\phi \in \Phi$ and local functions $f$, $complete(\phi, f) \Leftrightarrow f = \bla[\phi]$.

\proof
Assume $f$ = $bla[\phi]$. Then, by lemma \ref{blaconsequence}, we have that $\phi$ is a complete specification for $f$.

For the converse, assume $complete(\phi, f)$. We shall show that for any $\sigma \in \Sigma$, $f(\sigma) = \bla[\phi](\sigma)$. 

{\bf case 1: $f(\sigma) = \top$}. If $\bla[\phi](\sigma) \neq \top$, then $\bla[\phi] \models (\{\sigma\}, \bla[\phi](\sigma))$. This means that 
$\phi \models (\{\sigma\}, \bla[\phi](\sigma))$ (by lemma \ref{blaconsequence}), and so $f \models (\{\sigma\}, \bla[\phi](\sigma))$, but this is a contradiction. Therefore, $\bla[\phi](\sigma) = \top$  

{\bf case 2: $\bla[\phi](\sigma) = \top$}. If $f(\sigma) \neq \top$, then $f \models (\{\sigma\}, f(\sigma))$. This means that 
$\phi \models (\{\sigma\}, f(\sigma))$, and so $\bla[\phi] \models (\{\sigma\}, f(\sigma))$, but this is a contradiction. Therefore, $f(\sigma) = \top$  

{\bf case 3: $\bla[\phi](\sigma) \neq \top$ and $f(\sigma) \neq \top$}. We have 
\[\begin{array}{cl}
&f \models (\{\sigma\}, f(\sigma)) \\
\Rightarrow & \bla[\phi] \models (\{\sigma\}, f(\sigma)) \\
\Rightarrow & \bla[\phi](\sigma) \sqsubseteq f(\sigma)
\\\\
&\bla[\phi] \models (\{\sigma\}, \bla[\phi](\sigma)) \\
\Rightarrow & f \models (\{\sigma\}, \bla[\phi](\sigma)) \\
\Rightarrow & f(\sigma) \sqsubseteq \bla[\phi](\sigma)
\end{array}\]

Therefore $f(\sigma) = \bla[\phi](\sigma)$
\qed
\end{proposition}
Any specification is therefore only complete for a unique local function, which is its best local action. However, a local function may have lots of complete specifications. For example, if $\phi$ is a complete specification for $f$ and $(p, q) \in \phi$, then $\phi \cup \{(p, q')\}$ is also complete for $f$ if $q \subseteq q'$. For this reason it will be useful to have a canonical form for specifications.  

\begin{defi}[Canonicalisation]
The {\bf canonicalisation} of a specification $\phi$ is defined as $\phi_{can} = \{ (\{\sigma\}, \bla[\phi](\sigma)) \mid \sigma \in D(\phi) \} $. A specification is in {\bf canonical} form if it is equal to its canonicalisation. Let $\Phi_{can(f)}$ denote the set of all canonical complete specifications of $f$.
\end{defi}

Notice that a given local function does not necessarily have a \emph{unique} canonical complete specification. For example, both $\{(\{u\}, \{u\})\}$ and $\{(\{u\}, \{u\}), (\{\sigma\}, \{\sigma\})\}$, for some $\sigma \in \Sigma$, are canonical complete specifications for the identity function.

\begin{proposition}
\label{canproposition}
For any specification $\phi$, we have $\phi \Dashv \vDash \phi_{can}$.

\proof
We first show $\phi \vDash \phi_{can}$. For any $(p, q) \in \phi_{can}$, $(p, q)$ is of the form $(\{\sigma\}, \bla[\phi](\sigma))$ for some $\sigma \in D(\phi)$. So
we have $\bla[\phi] \models (p, q)$, and so $\phi \models (p, q)$ by lemma \ref{blaconsequence}.

We now show $\phi_{can} \vDash \phi$. For any $(p, q) \in \phi$, we have $\bla[\phi] \models (p, q)$. So for all 
$\sigma \in p$, $\bla[\phi](\sigma) \sqsubseteq q$, which implies 
$$ \displaystyle\bigsqcup_{\sigma \in p} \bla[\phi](\sigma) \sqsubseteq q \quad (*) $$
Now we have the following derivation:
\[
\infer
{
	(p, q)
}
{
	\infer
	{
		(\displaystyle\bigsqcup_{\sigma \in p} \{\sigma\}, \displaystyle\bigsqcup_{\sigma \in p} \bla[\phi](\sigma))  
	}
	{	
		\infer
  		{
  			(\{\sigma\}, \bla[\phi](\sigma)) \quad \mbox{\scriptsize{for all $\sigma \in p$}}
  		}
  		{
  			\phi_{can}
		}  						
	}
}
\]
The last step is by $(*)$ and consequence. So we have $\phi_{can} \vdash \phi$, and by soundness $\phi_{can} \models \phi$. 
\qed
\end{proposition}
Thus, the canonicalisation of a specification is logically equivalent to the specification. The following corollary shows that all complete specifications that have the same
domain have a unique canonical form, and specifications of different
domains have different canonical forms. 
\begin{cor}
\label{candomainisomorphism}
$\Phi_{can(f)}$ is isomorphic to the quotient set $\Phi_{comp(f)}/\cong_{D}$, under the isomorphism that maps $[\phi]_{\cong_{D}}$ to $\phi_{can}$, for every $\phi \in \Phi_{comp(f)}$.
\proof
By proposition \ref{blaiffcomplete}, all complete specifications for $f$ have the same best local action, which is $f$ itself. So by the definition of canonicalisation, it can be seen that complete specifications with different domains have different canonicalisations, and complete specifications with the same domain have the same canonicalisation. This shows that the mapping is well-defined and injective. Every canonical complete specification $\phi$ is also complete, and $[\phi]_{\cong_{D}}$ maps to $\phi_{can} = \phi$, so the mapping is surjective.  
\qed
\end{cor}
\begin{defi}[Small and Big specifications]
\label{bigsmallspec}
$\phi$ is a {\bf small specification} for $f$ iff $\phi \in \Phi_{comp(f)}$ and there is no $\psi \in \Phi_{comp(f)}$ such that $D(\psi) \sqsubset 
D(\phi)$. A {\bf big specification} is defined similarly. 
\end{defi}
{\em Small} and {\em big} specifications are thus the specifications with the smallest and biggest domains respectively. The question is if/when small and big specifications exist. The following result shows that a canonical big specification exists for every local function. 
\begin{proposition}[Big Specification]
\label{bigspec}
For any local function $f$, the canonical big specification for $f$ is given by $\phi_{big(f)} = \{ (\{\sigma\}, f(\sigma)) \mid f(\sigma) \sqsubset \top \}$.

\proof
$f \models \phi_{big(f)}$ is trivial to check. To show $complete(\phi_{big(f)}, f)$, assume $f \models (p, q)$ for some $p, q \in \mathcal{P}(\Sigma)$. Note that, for 
any $\sigma \in p$, $f(\sigma) \sqsubseteq q$ and so $\displaystyle\bigsqcup_{\sigma \in p} f(\sigma) \sqsubseteq q$. We then have the derivation
\[
\infer
{
	(p, q)
}
{
	\infer
	{
		(\displaystyle\bigsqcup_{\sigma \in p} \{\sigma\}, \displaystyle\bigsqcup_{\sigma \in p} f(\sigma))  
	}
	{	
		\infer
  		{
  			(\{\sigma\}, f(\sigma)) \quad \mbox{\scriptsize{for all $f(\sigma) \sqsubset \top$}}
  		}
  		{
  			\phi_{big(f)}
		}  						
	}
}
\]
By soundness we get $\phi_{big(f)} \models (p, q)$.
$\phi_{big(f)}$ has the biggest domain because $f$ would fault
on any element not included in $\phi_{big(f)}$.
\qed
\end{proposition}

The notion of a small specification has until now been  used in an informal sense in local reasoning papers \cite{ORY01,BCOP05,CGZ05} as specifications that completely specify the behaviour of an update command by only describing the command's  behaviour on the part of the resource that it affects. Although these papers present examples of such specifications for specific commands, the notion has so far not received a formal treatment in the general case. The question of the existence of small specifications is strongly related to the concept of footprints, since finding a small specification is about finding a complete specification with the smallest possible domain, and therefore enquiring about which elements of $\Sigma$ are essential and sufficient for a complete specification. This requires a formal characterisation of the footprint notion, which we shall now present.

\section{Footprints}
\label{sec:footprints}

In the introduction we discussed how the \emph{AD} program demonstrates that the footprints of a local function do not correspond simply to the smallest safe states, as these states alone do not always yield complete specifications. In this section we introduce the definition of footprint that does yield complete specifications. In order to understand what the footprint of a local function should be, we begin by analysing the definition of locality. Recall that the definition of locality (definition \ref{def:localaction}) says that the action on a certain state $\sigma_1$ imposes a \emph{limit} on the action on a bigger state $\sigma_2\bullet\sigma_1$. This limit is $\{\sigma_2\} * f(\sigma_1)$, as we have $f(\sigma_2\bullet \sigma_1) \sqsubseteq \{\sigma_2\} * f(\sigma_1)$. 

Another way of viewing this definition is that  for any state $\sigma$, the action of the function on that state has to be within the limit imposed by \emph{every} substate $\sigma'$ of  $\sigma$, that is, $f(\sigma) \sqsubseteq \{\sigma - \sigma'\} * f(\sigma')$. In the case where $\sigma' = \sigma$, this condition is trivially satisfied for any function (local or non-local). The distinguishing characteristic of local functions is that this condition is also satisfied by every strict substate of $\sigma$, and thus we have
\[f(\sigma) \sqsubseteq \displaystyle\bigsqcap_{\sigma' \prec \sigma} \{\sigma - \sigma'\} * f(\sigma')\]
We define this overall constraint imposed on $\sigma$ by all of its strict substates as the \emph{local limit} of $f$ on $\sigma$, and show that the locality definition is equivalent to satisfying the local limit constraint.  
\begin{defi}[Local limit]
For a local function $f$ on $\Sigma$ and $\sigma \in \Sigma$, the {\bf local limit} of $f$ on $\sigma$ is defined as
\[L_f(\sigma) =  \displaystyle\bigsqcap_{\sigma' \prec \sigma} \{\sigma - \sigma'\} * f(\sigma')\]
\end{defi}
\begin{proposition}
\label{locallimitproposition}$f \mbox{ is local} \quad \Leftrightarrow \quad f(\sigma) \sqsubseteq L_f(\sigma) \quad \mbox{for all $\sigma \in \Sigma$}$

\proof
Assume $f$ is local. So for any $\sigma$, for every $\sigma' \prec \sigma$, $f(\sigma) \sqsubseteq \{\sigma - \sigma'\} * f(\sigma')$. $f(\sigma)$ is 
therefore smaller than the intersection of all these sets, which is $L_f(\sigma)$. 

For the converse, assume the rhs and that $\sigma_1 \bullet \sigma_2$ is defined. If $\sigma_1 = u$ then 
$f(\sigma_1 \bullet \sigma_2) \sqsubseteq \{\sigma_1\} * f(\sigma_2)$ and we are done. Otherwise,  $\sigma_2 \prec \sigma_1 \bullet \sigma_2$ and we have 
$f(\sigma_1 \bullet \sigma_2) \sqsubseteq L_f(\sigma_1 \bullet \sigma_2) \sqsubseteq \{\sigma_1\} * f(\sigma_2)$.
\qed
\end{proposition}

Thus for any local function $f$ acting on a certain state $\sigma$, the local limit determines a \emph{smallest upper bound} on the possible outcomes on $\sigma$, based on the outcomes on all smaller states. If this smallest upper bound does correspond exactly to the set of all possible outcomes on $\sigma$, then $\sigma$ is `large enough' that just the action of $f$ on smaller states and the locality of $f$ determines the complete behaviour of $f$ on $\sigma$. In this case we will not think of $\sigma$ as a footprint of $f$, as smaller states are sufficient to determine the action of $f$ on $\sigma$. With this observation, we define footprints as those states on which the outcomes cannot be determined only by the smaller states, that is, the set of outcomes is a \emph{strict} subset of the local limit. 

\begin{defi}[Footprint]
\label{def:footprint}
For a local function $f$ and $\sigma \in \Sigma$, $\sigma$ is a footprint of $f$, written $F_f(\sigma)$, iff $f(\sigma) \sqsubset L_f(\sigma)$. We denote the set of footprints of $f$ by $F(f)$. 
\end{defi}

Note that an element $\sigma$ is therefore not a footprint if and only if the action of $f$ on $\sigma$ is at the local limit, that is $f(\sigma) = L_f(\sigma)$. 

\begin{lemma}
\label{minstates}
For any local function $f$, the smallest safe states of $f$ are footprints of $f$. 
\proof
Let $\sigma$ be a smallest safe state for $f$. Then for any $\sigma' \prec \sigma$, $f(\sigma') = \top$. Therefore $L_f(\sigma) = \top$ and so $f(\sigma) \sqsubset L_f(\sigma)$.
\qed
\end{lemma}
 
However, the smallest safe states are not always the \emph{only} footprints. An example is the \emph{AD} command discussed in the introduction. The empty heap is a footprint as it is the smallest safe heap, but the heap cell $l \points v$ is also a footprint. 

\begin{exa}[Dispose]
\label{dispose}
The footprints of the $dispose[l]$ command in the plain heap model (example \ref{locsepexamples}.1) are the cells at location $l$. We check this by considering the following cases
\begin{enumerate}[(1)]

\item The empty heap, $u_H$, is not a footprint since $L_{dispose[l]}(u_H) = \top = dispose[l](u_H)$
\item{Every cell $l \points v$ for some $v$ is a footprint}
\[
\begin{array}{l}
L_{dispose[l]}(l \points v) =  \{l \points v\} * dispose[l](u_H) =  \{l \points v\} * \top =  \top\\
dispose[l](l \points v) = \{u_H\} \sqsubset L_{dispose[l]}(l \points v)
\end{array}
\]
\item Every state $\sigma$ such that $\sigma \succ (l \points v)$ for some $v$ is not a footprint
\[L_{dispose[l]}(\sigma) \sqsubseteq  \{\sigma - (l \points v)\} * dispose[l](l \points v) =  \{\sigma - (l \points v)\} =  dispose[l](\sigma)\]
By proposition \ref{locallimitproposition}, we have $L_{dispose[l]}(\sigma) =  dispose[l](\sigma)$. The intuition is that $\sigma$ does not characterise any `new'  behaviour of the function: its action on $\sigma$ is just a consequence of its action on the cells at location $l$ and the locality property of the function.
\item Every state $\sigma$ such that $\sigma \not \succ (l \points v)$ for some $v$ is not a footprint
\[L_{dispose[l]}(\sigma) \sqsubseteq  \{\sigma\} * dispose[l](u_H) =  \{\sigma\} * \top = \top =  dispose[l](\sigma)\]
Again by proposition \ref{locallimitproposition}, $L_{dispose[l]}(\sigma) =  dispose[l](\sigma)$. 
\end{enumerate}
\end{exa}

\begin{exa}[{\em AD} command]
\label{ad}
The \emph{AD} (Allocate-Deallocate) command was defined on the heap and stack model in example \ref{locsepexamples}.2. We have the following cases for $\sigma$.
\begin{enumerate}[(1)]
\item {$\sigma \not \succeq x \points v_1$ for some $v_1$ is not a footprint, since $L_{AD}(\sigma) = \top = AD(\sigma)$}.

\item {$\sigma =  x \points v_1$ for some $v_1$ is a footprint since $L_{AD}(\sigma) = \top$ (by case (1)) and
$AD(\sigma) = \{x \points w \mid w \in L\} \sqsubset L_{AD}(\sigma)$}.

\item {$\sigma = l \points v_1 \bullet x \points v_2$ for some $l, v_1, v_2$ is a footprint.}
\[
\begin{array}{ll}
L_{AD}(\sigma) &=  \{l \points v_1\} * AD(x \points v_2)\\
&\quad \mbox{(AD faults on all other elements strictly smaller than $\sigma$)}\\
&=  \{l \points v_1\} * \{ x \points w \mid w \in L\}\\
&=  \{l \points v_1 \bullet x \points w \mid w \in L\}\\\\
AD(\sigma) &= \{l \points v_1\bullet x \points w \mid w \in L, w \neq l\} \sqsubset L_{AD}(\sigma)
\end{array}
\]

\item {$\sigma = h\bullet x \points v_1$ for some $v_1$, and where $|loc(h)| > 1$, is not a footprint}.
\[\eqalign{
L_{AD}(\sigma) &\sqsubseteq  \displaystyle\bigsqcap_{h \succ l \;\points\; v} \{(h - l \points v\} * AD(l \points v\bullet x \points v_1) \cr
&= \{h\bullet x \points w \mid w \not \in loc(h)\} = AD(\sigma)}
\]
By proposition \ref{locallimitproposition}, we get $L_{AD}(\sigma) = AD(\sigma)$.
\end{enumerate}
\end{exa}

\noindent Our footprint definition therefore works properly for these
specific examples. Now we give the formal general result which
captures the underlying intuition of local reasoning, that the
footprints of a local function are the only essential elements for a
complete specification of the function.

\begin{thm}[Essentiality]
\label{essentialitytheorem}
The footprints of a local function are the essential domain elements for any complete specification of that function, that is,
\[F_f(\sigma) \quad \Leftrightarrow \quad \forall \phi \in \Phi_{comp(f)}.\: \sigma \in D(\phi)\]
\end{thm}

\proof
Assume some fixed $f$ and $\sigma$. We establish the following equivalent statement : 
    $$  \neg F_f(\sigma) \quad \Leftrightarrow \quad   \exists \phi\in \Phi_{comp(f)}.\: \sigma \not \in D(\phi) $$
We first show the right to left implication. So assume $\phi$ is
     a complete specification of $f$ such that $\sigma\not\in D(\phi)$. Since $complete(\phi, f)$, by proposition \ref{blaiffcomplete}, we have $f = \bla[\phi]$. So 
\[f(\sigma) = \bigsqcap_{\sigma_1 \preceq \sigma, \sigma_1\in p, (p, q) \in \phi} \{\sigma - \sigma_1\} * q \]
Now for any set $\{\sigma - \sigma_1\} * q$ in the above intersection, we have that $\sigma_1 \in p$, and $(p, q) \in \phi$ for
some $p$. Since $\sigma_1 \in p$, we have $f(\sigma_1) \sqsubseteq q$, and therefore $\{\sigma - \sigma_1\} * f(\sigma_1) \sqsubseteq \{\sigma - \sigma_1\} * q$.
Also, $\sigma_1 \neq \sigma$, because otherwise we would have $\sigma \in p$, which would contradict the assumption that $\sigma \notin D(\phi)$. So $\sigma_1 \prec \sigma$ and we have
$$ L_f(\sigma) \sqsubseteq \{\sigma - \sigma_1\} * f(\sigma_1) \sqsubseteq \{\sigma - \sigma_1\} * q $$
So the local limit is smaller than each set $\{\sigma - \sigma_1\} * q$ in the intersection, and therefore it is smaller than the intersection itself: $ L_f(\sigma) \sqsubseteq f(\sigma)$. We know from proposition \ref{locallimitproposition} that $f(\sigma) \sqsubseteq L_f(\sigma)$, so we get $f(\sigma) = L_f(\sigma)$ and therefore $\neg F_f(\sigma)$.

We now show the left to right implication. Assume that $\sigma$ is not a footprint of $f$. We shall use the big specification, $\phi_{big(f)}$, to construct a complete specification of $f$ which does not contain $\sigma$ in its domain. If $f(\sigma) = \top$ then the big specification itself is such a specification, and we are done. Otherwise assume $f(\sigma) \sqsubset \top$. Let $\phi = \phi_{big(f)} / \{(\{\sigma\}, f(\sigma))\}$. It can be seen that $\sigma \notin D(\phi)$. Now we need to show that $\phi$ is complete for $f$. For this it is 
sufficient to show $\phi \dashv \vdash \phi_{big(f)}$ because we know that $\phi_{big(f)}$ is complete for $f$. The right to left direction, $\phi \dashv \phi_{big(f)}$, is trivial.

For $\phi \vdash \phi_{big(f)}$, we just need to show $\phi \vdash (\{\sigma\}, f(\sigma))$. We have the following derivation: 
\[
\infer
{
	(\{\sigma\}, L_f(\sigma))
}
{
	\infer
	{
		(\{\sigma\}, \displaystyle\bigsqcap_{\sigma' \prec \sigma, f(\sigma') \sqsubset \top} \{\sigma - \sigma'\} * f(\sigma'))  
	}
	{	
		\infer
  		{
  			(\{\sigma - \sigma'\} * \{\sigma'\}, \{\sigma - \sigma'\} * f(\sigma')) \quad \mbox{\scriptsize{for all $\sigma' \prec \sigma$, $f(\sigma') \sqsubset \top$}}
  		}
  		{
  			\infer
  			{
  				(\{\sigma'\}, f(\sigma')) \quad \mbox{\scriptsize{for all $\sigma' \prec \sigma$, $f(\sigma') \sqsubset \top$}}
  			}
  			{
  				\phi
			}
		}	  						
	}
}
\]
The intersection rule can be safely applied as there is at least one $\sigma' \prec \sigma$ such that $f(\sigma') \sqsubset \top$. This is because $f(\sigma) \sqsubset \top$, so if there were no such $\sigma'$ then $\sigma$ would be a footprint, which is a contradiction. Note that the last step uses the fact that 
$$\displaystyle\bigsqcap_{\sigma' \prec \sigma, f(\sigma') \sqsubset \top} \{\sigma - \sigma'\} * f(\sigma') = 
\displaystyle\bigsqcap_{\sigma' \prec \sigma} \{\sigma - \sigma'\} * f(\sigma') = L_f(\sigma)$$
because adding the top element to an intersection does not change its value. Since $\sigma$ is not a footprint, $f(\sigma) = L_f(\sigma)$, and so $\phi \vdash (\{\sigma\}, f(\sigma))$. 
\qed

\section{Sufficiency and Small Specifications}

We know that the footprints are the only elements that are {\em essential} for a complete specification of a local function in the sense that every complete specification must include them. Now we ask when a set of elements is {\em sufficient} for a complete specification of a local function, in the sense that there exists a complete specification of the function that only includes these elements. In particular, we wish to know if the footprints alone are sufficient. To study this, we begin by identifying the notion of the \emph{basis} of a local function.

\subsection{Bases}

In the last section we defined the local limit of a function $f$ on a state $\sigma$ as the constraint imposed on $f$ by all the strict substates of $\sigma$. This was used to identify the footprints as those states on which the action of $f$ cannot be determined by just its action on the smaller states. We are now addressing the question of when a set of states is {\em sufficient} to determine the behaviour of $f$ on any state. We shall do this by identifying a fixed set of states, which we call a {\em basis} for $f$, such that the action of $f$ on any state $\sigma$  can be determined by just the substates of $\sigma$ taken from this set (rather than all the strict substates of $\sigma$). Thus we first generalise the  local limit definition to consider the constraint imposed by only the substates taken from a given set. 

\begin{defi}[Local limit imposed by a set]
For a subset $A$ of a separation algebra $\Sigma$, the {\bf local limit} imposed by $A$ on the action of $f$ on $\sigma$ is defined by
\[ L_{A, f}(\sigma) =  \displaystyle\bigsqcap_{\sigma' \preceq \sigma, \sigma' \in A} \{\sigma - \sigma'\} * f(\sigma')\]
\end{defi}
Sometimes, the local limit imposed by $A$ is enough to completely determine $f$. In this case, we call $A$ a \emph {basis} for $f$. 
\begin{defi}[Basis] $A \sqsubseteq \Sigma$ is a {\bf basis} for $f$, written $basis(A, f)$, iff $L_{A, f} = f$.
\end{defi}
This means that, when given the action of $f$ on elements in A alone, we can determine the action of $f$ on any element in $\Sigma$ by just using the locality property of $f$. Every local function has at least one basis, namely the trivial basis $\Sigma$ itself. We next show the correspondence between the bases and complete specifications of a local function. 
\begin{lemma}
\label{basisspeclemma}
Let $\phi_{A, f} = \{ (\{\sigma\}, f(\sigma)) \mid \sigma \in A, f(\sigma) \sqsubset \top\}$. Then we have $basis(A, f) \Leftrightarrow complete(\phi_{A, f}, f)$.
\proof
We have $L_{A, f} = \bla[\phi_{A, f}]$ by definition. The result follows by proposition \ref{blaiffcomplete} and the definition of basis.
\qed
\end{lemma}
For every canonical complete specification $\phi \in \Phi_{can(f)}$, we have $\phi = \phi_{D(\phi), f}$. By the previous lemma it follows that $D(\phi)$ forms a basis for $f$. The lemma therefore shows that every basis determines a complete canonical specification, and vice versa. This correspondence also carries over to all complete specifications for $f$ by the fact that every domain-equivalent class of complete specifications for $f$ is represented by the canonical complete specification with that domain (corollary \ref{candomainisomorphism}). By the essentiality of footprints (theorem \ref{essentialitytheorem}), it follows that the footprints are present in every basis of a local function.
\begin{lemma}
\label{footprintbasislemma}
The footprints of $f$ are included in every basis of f. 
\proof
Every basis $A$ of $f$ determines a complete specification for $f$ the domain of which is a subset of $A$. By the essentiality theorem (\ref{essentialitytheorem}), the domain includes the footprints.
\qed
\end{lemma}

The question of sufficiency is about how small the basis can get. Given a local function, we wish to know if it has a smallest basis.

\subsection{Well-founded Resource}

We know that every basis must contain the footprints. Thus if the footprints alone form a basis, then the function will have a \emph {smallest} complete specification whose domain are just the footprints. We find that, for well-founded resource models, this is indeed the case.  
\begin{thm}[Sufficiency I]
\label{sufficiencytheorem}
If a separation algebra $\Sigma$ is well-founded under the $\preceq$ relation, then the footprints of any local function form a basis for it, that is, $f = L_{F(f), f}$.
\end{thm}

\proof
Assume that $\Sigma$ is well-founded under $\preceq$. We shall show by
induction that $ f(\sigma) = L_{F(f), f}(\sigma)$ for all $\sigma \in
\Sigma$. The induction hypothesis is that, for all $\sigma' \prec
\sigma$, $ f(\sigma') = L_{F(f), f}(\sigma')$

{\bf case 1:} Assume $\sigma$ is a footprint of $f$. We have
$f(\sigma) = \{u\} * f(\sigma)$ is in the intersection in the
definition of $L_{F(f), f}(\sigma)$, and so $L_{F(f), f}(\sigma)
\sqsubseteq f(\sigma)$. We have by locality that $f(\sigma)
\sqsubseteq L_{F(f), f}(\sigma)$, and so $f(\sigma) = L_{F(f),
  f}(\sigma)$.

{\bf case 2:} Assume $\sigma$ is not a footprint of $f$. We have
\[\eqalign{
  f(\sigma) 
&=L_f(\sigma) \quad \mbox{\emph{(because $\sigma$ is not a footprint of f)}}\cr
&=\bigsqcap_{\sigma' \prec \sigma} \{\sigma - \sigma'\} * f(\sigma')\cr
&=\bigsqcap_{\sigma' \prec \sigma} \big(\{\sigma - \sigma'\}*\bigsqcap_{\sigma'' \preceq \sigma', F_f(\sigma'')} \{\sigma' - \sigma''\} * f(\sigma'')\big) \quad \mbox{\emph {(by the induction hypothesis)}} \cr
&=\bigsqcap_{\sigma' \prec \sigma, \sigma'' \preceq \sigma', F_f(\sigma'')} \{\sigma - \sigma'\} * \{\sigma' - \sigma''\} * f(\sigma'') \quad \mbox{\emph{(by the precision of $\{\sigma - \sigma'\}$)}}\cr
&=\bigsqcap_{\sigma'' \prec \sigma, F_f(\sigma'')} \{\sigma - \sigma''\} * f(\sigma'')\cr
&=\bigsqcap_{\sigma'' \preceq \sigma, F_f(\sigma'')} \{\sigma - \sigma''\} * f(\sigma'') \quad \mbox{\emph{(because $\sigma$ is not a footprint of f)}}\cr
&=  L_{F(f), f}(\sigma)\rlap{\hbox to319 pt{\hfill}\qEd}}
\]
In section \ref{completespecifications}, the notions of big and small
specifications were introduced (definition \ref{bigsmallspec}), and
the existence of a big specification was shown (proposition
\ref{bigspec}). We are now in a position to show the existence of the
small specification for well-founded resource. If $\Sigma$ is
well-founded, then every local function has a small specification
whose domain is the footprints of the function.
\begin{cor}[Small specification]
\label{smallspeccorollary}
For well-founded separation algebras, every local function has a small specification given by $\phi_{F(f), f}$.
\proof
$\phi_{F(f), f}$ is complete by theorem \ref{sufficiencytheorem} and lemma \ref{basisspeclemma}. It has the smallest domain by the essentiality theorem.  
\qed
\end{cor}

Thus, for well-founded resource, the footprints are always essential and sufficient, and specifications need not consider any other elements. In practice, small specifications may not always be in canonical form even though they always have the same domain as the canonical form. For example, the heap dispose command can have the specification $\{(\{l \points v \mid v \in Val\}, \{u_H\})\}$ rather than the canonical one given by $\{(\{l \points v\}, \{u_H\}) \mid v \in Val\}$. 

In practical examples it is usually the case that resource is well-founded.  A notable exception is the fractional permissions model \cite{BCOP05} in which the resource includes `permissions to access', which can be indefinitely divided. We next investigate the non-well-founded case.

\subsection{Non-well-founded Resource}

If a separation algebra is non-well-founded under the $\preceq$ relation, then there is some infinite descending chain of elements $\sigma_1 \succ \sigma_2 \succ \sigma_3 ...$. From a resource-oriented point of view, there are two distinct ways in which this could happen. One way is when it is possible to remove non-empty pieces of resource from a state indefinitely, as in the separation algebra of non-negative real numbers under addition. In this case any infinite descending chain does not have more than one occurrence of any element. Another way is when an infinite chain may exist because of repeated occurrences of some elements. This happens when there is \emph{negativity} present in the resource: some elements have inverses in the sense that adding two non-unit elements together may give the unit.  An example is the separation algebra of integers under addition, where $1 + (-1) = 0$, so adding -1 to 1 is like adding negative resource. Also, since $1 = 0 + 1$, we have that  $1 \succ 0 \succ 1 ...$ forms an infinite chain. 

\begin{defi}[Negativity]
A separation algebra $\Sigma$ has {\bf negativity} iff there exists a non-unit element $\sigma \in \Sigma$ that has an inverse; that is, $\sigma \neq u$ and $\sigma \bullet \sigma' = u$ for some $\sigma' \in \Sigma$. We say that $\Sigma$ is {\bf non-negative} if no such element exists. 
\end{defi}

All separation algebras with negativity are non-well-founded because, for elements $\sigma$ and $\sigma'$ such that $\sigma\bullet\sigma' = u$, the set $\{\sigma, u\}$ forms an infinite descending chain (there is no least element). All well-founded models are therefore non-negative. For the general non-negative case, we find that either the footprints form a basis, or there is no smallest basis. 

\begin{thm}[Sufficiceny II]
\label{nonnegativetheorem}
If $\Sigma$ is non-negative then, for any local $f$, either the footprints form a smallest basis or there is no smallest basis for f.
\end{thm}

\proof
Let $A$ be a basis for $f$ (we know there is at least one, which is
the trivial basis $\Sigma$ itself). If $A$ is the set of footprints
then we are done. So assume $A$ contains some non-footprint $\mu$. We
shall show that there exists a smaller basis for $f$, which is
$A/\{\mu\}$. So it suffices to show $f(\sigma) = L_{A/\{\mu\},
  f}(\sigma)$ for all $\sigma \in \Sigma$. 


  {\bf case 1:} $\mu \not \preceq \sigma$.  We have 
\[f(\sigma)
 =L_{A, f}(\sigma)
 =\bigsqcap_{\sigma'\preceq\sigma,\sigma'\in A}\{\sigma-\sigma'\}*f(\sigma')
 =\bigsqcap_{\sigma'\preceq\sigma,\sigma'\in A/\{\mu\}}
\{\sigma-\sigma'\}*f(\sigma')= L_{A/\{\mu\}, f}(\sigma)
\]
  as desired

  {\bf case 2:} $\mu \preceq \sigma$.  This implies
\[f(\sigma)=\Bigl(\bigsqcap_{\sigma'\preceq\sigma,\sigma'\in
  A/\{\mu\}}\{\sigma-\sigma'\}*f(\sigma')\Bigr)\enspace\sqcap\enspace
  (\{\sigma-\mu\}*f(\mu))
\]
It remains to show that the right hand side of this
intersection  contains the left hand side:
\[\eqalign{
\{\sigma - \mu\} * f(\mu)
&=\{\sigma - \mu\} * L_f(\mu) 
  \quad\mbox{{(because $\mu$ is not a footprint of f)}} \cr
&=\{\sigma-\mu\}*\bigsqcap_{\sigma' \prec \mu}\{\mu-\sigma'\}*f(\sigma')\cr
&=\{\sigma-\mu\}*\displaystyle\bigsqcap_{\sigma'\prec\mu}\big(\{\mu-\sigma'\}*
  \bigsqcap_{\sigma''\preceq\sigma',\sigma''\in A/\{\mu\}}\{\sigma'-\sigma''\}*
   f(\sigma'') \big)\cr
&\enspace\quad\mbox{{(case 1 applies because $\Sigma$ is non-negative, 
  so $\sigma' \prec \mu \Rightarrow \mu \not \preceq \sigma'$)}} \cr
&=\bigsqcap_{\sigma' \prec \mu}\bigsqcap_{\sigma'' \preceq \sigma', \sigma'' \in A/\{\mu\}} \{\sigma - \mu\} * \{\mu - \sigma'\} * \{\sigma' - \sigma''\} * f(\sigma'') \quad\mbox{{(by precision)}} \cr
&=\bigsqcap_{\sigma' \prec \mu}\bigsqcap_{\sigma'' \preceq \sigma', \sigma'' \in A/\{\mu\}} \{\sigma - \sigma''\} * f(\sigma'') \cr
&=\bigsqcap_{\sigma'' \prec \mu, \sigma'' \in A/\{\mu\}} \{\sigma - \sigma''\} * f(\sigma'') \cr
&\sqsupseteq\bigsqcap_{\sigma'' \preceq \sigma, \sigma'' \in
     A/\{\mu\}} \{\sigma - \sigma''\} * f(\sigma'')
\rlap{\hbox to194 pt{\hfill}\qEd}\cr}
\]
\begin{cor}[Small Specification]
\label{nonnegativecorollary}
If $\Sigma$ is non-negative, then every local function either has a small specification given by $\phi_{F(f), f}$ or there is no smallest complete specification for that function.
\end{cor}

\begin{exa}[Permissions]
\label{permissions}
The fractional permissions model \cite{BCOP05} is non-well-founded and non-negative. It can be represented by the separation algebra $\mathit{HPerm} = L \rightharpoonup_{\fin} Val\times P$ where $L$ and $Val$ are as in example \ref{locsepexamples}, and $P$ is the interval (0, 1] of rational numbers. Elements of $P$ represent `permissions' to access a heap cell. A permission of 1 for a cell means both read and write access, while any permission less than 1 is read-only access. The operator $\bullet$ joins disjoint heaps and adds the permissions together for any cells that are present in both heaps only if the resulting permission for each heap cell does not exceed 1; the operation is undefined otherwise. In this case, the write function that updates the value at a location requires a permission of at least 1 and faults on any smaller permission. It therefore has a small specification with precondition being the cell with permission 1. The read function, however, can execute safely on any positive permission, no matter how small. Thus, this function can be completely specified with a specification that has a precondition given by the cell with permission $z$, for all $0 < z \leq 1$. However, this is not a \emph{smallest} specification, as a smaller one can be given by further restricting $0 < z \leq 0.5$. We can therefore always find a smaller specification by reducing the value of $z$ but keeping it positive.   
\end{exa}

For resource with negativity, we find that it is possible to have small specifications that include non-essential elements (which by theorem \ref{essentialitytheorem} are not footprints). These elements are non-essential in the sense that complete specifications exist that do not include them, but there is no complete specification that includes only essential elements.

\begin{exa}[Integers]
\label{negativeresource}
An example of a model with negativity is the separation algebra of integers $({\mathbb{Z}}, +, 0)$. In this case there can be local functions which can have small specifications that contain non-footprints. Let $f:{\mathbb{Z}} \rightarrow \mathcal{P}({\mathbb{Z}})^\top$ be defined as $f(n) = \{n + c\}$ for some constant $c$, as in example \ref{locsepexamples}. $f$ is local, but it has no footprints. This is because for any $n$, $f(n) = 1 + f(n - 1)$, and so $n$ is not a footprint of $f$. However, $f$ does have small specifications, for example, $\{(\{0\}, \{c\})\}$, $\{(\{5\}, \{5 + c\})\}$, or indeed $\{(\{n\}, \{n + c\})\}$ for any $n \in {\mathbb{Z}}$. So although every element is non-essential, some element is required to give a complete specification.
\end{exa}

\section{Regaining Safety Footprints}
\label{safety} 

In the introduction we discussed how the notion of footprints as the smallest safe states - the \emph{safety footprint}- is inadequate for giving complete specifications, as illustrated by the \emph{AD} example. For this reason, so far in this paper we have investigated the general notion of footprint for arbitrary local functions on arbitrary separation algebras. Equipped with this general theory, we now investigate how the regaining of safety footprints may be achieved with different resource modelling choices. We start by presenting an alternative model of RAM, based on an investigation of why the $AD$ phenomenon occurs in the standard model. We then demonstrate that the footprints of the $AD$ command in this new model do correspond to the safety footprints. In the final section we identify, for arbitrary separation algebras, a condition on local functions which guarantees the equivalence of the safety footprint and the actual footprint. We then show that if this condition is met by all the primitive commands of a programming language then the safety footprints are regained for every program in the language, and finally show that this is indeed the case in our new RAM model. 

\subsection{An alternative model}
\label{altmodel}

In this section we explore an alternative heap model in which the safety footprints do correspond to the actual footprints. We begin by taking a closer look at why the \emph{AD} anomaly occurs in the standard heap and stack model described in example \ref{locsepexamples}.2. Consider an application of the allocation command in this model:
\[
\mathit{new[x]}(42 \mapsto v \bullet x \mapsto w) = \{42 \mapsto v \bullet x \mapsto l  \bullet l \mapsto r \mid l \in L\backslash \{42\}, r \in Val\}
\]

The intuition of locality is that the initial state $42 \mapsto v \bullet x \mapsto w$ is only describing a local region of the heap and the stack, rather than the whole global state. In this case it says that the address 42 is initially allocated, and the definition of the allocation command is that the resulting state will have a new cell, the address of which can be anything other than 42. However, we notice that the initial state is in fact not just describing only its local region of the heap. It does state that 42 is allocated, but it also implicitly states a very global property: that \emph {all other addresses are not allocated}. This is why the allocation command can choose to allocate any location that is not 42. Thus in this model, every local state implicitly contains some global allocation information which is used by the allocation command. In contrast, a command such as mutate does not require this global `knowledge' of the allocation status of any other cell that it is not affecting. Now the global information of which cells are free {\em changes} as more resource is added to the initial state, so this can lead to program behaviour being sensitive to the addition of more resource to the initial state, and this sensitivity is apparant in the case of the \emph{AD} program. 

Based on this observation, we consider an alternative model. As before, a state $l \mapsto v$ will represent a local allocated region of the heap at address $l$ with value $v$. However, unlike before, this state will say nothing about the allocation status any locations other than $l$. This information about the allocation status of other locations will be represented explicitly in a \emph{free} set, which will contain every location that is not allocated in the \emph{global heap}. The model can be interpreted from an ownership point of view, where the free set is to be thought of as a unique, atomic piece of resource, ownership of which needs to be obtained by a command if it wants to do allocation or deallocation. An analogy is with the permissions model: a command that wants to read or write to a cell needs ownership of the appropriate permission on that cell. In the same way, in our new model, a command that wants to do allocation or deallocation needs to have ownership of the free set: the `permission' to see which cells are free in the global heap so that it can choose one of them to allocate, or update the free set with the address that it deallocates. On the other hand, commands that only read or write to cells shall not require ownership of the free set.

\begin{exa}[Heap model with free set]
\label{freesetmodel}
Formally, we work with a separation algebra $(H,\bullet,u_H)$. Let $L$, $Var$ and $Val$ be locations, variables and values, as before. States $h \in H$ are given by the grammar: 
$$h ::= u_H \mid l \points v \mid x \points v \mid F \mid h \bullet h$$
where $l \in L$, $v \in Val$, $x \in Var$ and $F \in \mathcal{P}(L)$. The operator $\bullet$ is undefined for states with overlapping locations or variables. Let $loc(h)$ and $var(h)$ be the set of locations and variables in state $h$ respectively. The set $F$ carries the information of which locations are free. Thus we allow at most one free set in a state, and the free set must be disjoint from all locations in the state. So $h \bullet F$ is only defined when $loc(h) \cap F = \emptyset$ and $h \neq h'\bullet F'$ for any $h'$ and $F'$. We assume $\bullet$ is associative and commutative with unit $u_H$.  

In this model, the allocation command requires ownership of the free set for safe execution, since it chooses the location to allocate from this set. It removes the chosen address from the free set as it allocates the cell. It is defined as
\[
new[x](h)
= \left\{ \begin{array}{ll}
\begin{array}{l}\{ h' \bullet x \points l \bullet l \points w \bullet F\backslash \{l\}\mid w \in Val, l \in F\} \end{array} &  \begin{array}{l}h = h'\bullet x \points v \bullet F \end{array} \\
\begin{array}{l}\top \end{array} & \begin{array}{l}\mbox{otherwise} \end{array}
\end{array}\right. 
\]
Note that the output states $h' \bullet x \points l \bullet l \points w \bullet F\backslash \{l\}$ are defined, since we have $l \not \in F\backslash\{l\}$ and the input state $h'\bullet x \points v \bullet F$ implies that $loc(h')$ is disjoint from $F\backslash\{l\}$. The deallocation command also requires the free set, as it updates the set with the address of the cell that it deletes:
\[
dispose[x](h)
= \left\{ \begin{array}{ll}
\begin{array}{l}\{ h' \bullet x \points l \bullet F \cup \{l\}\} \end{array} &  \begin{array}{l}h = h'\bullet x \points l \bullet l \points v \bullet F \end{array} \\
\begin{array}{l}\top \end{array} & \begin{array}{l}\mbox{otherwise} \end{array}
\end{array}\right. 
\]
Again, the output states are defined, since the input state implies that $loc(h') \cup \{l\}$ is disjoint from $F$, and so $loc(h')$ is disjoint from $F \cup \{l\}$. Notice that in this model, only the allocation and deallocation commands require ownership of the free set, since commands such as mutation and lookup are completely independent of the allocation status of other cells, and they are defined exactly as in example \ref{locsepexamples}.2:
\[\eqalign{
mutate[x,v](h)
&= \left\{ \begin{array}{ll}
\begin{array}{l}\{ h' \bullet x \points l \bullet l \points v \} \end{array} &  \begin{array}{l}h = h'\bullet x \points l\bullet l \points w \end{array} \\
\begin{array}{l}\top \end{array} & \begin{array}{l}\mbox{otherwise} \end{array}
\end{array}\right.\cr
lookup[x,y](h)
&= \left\{ \begin{array}{ll}
\begin{array}{l}\{ h' \bullet x \points l \bullet l \points v \bullet y \points v \} \end{array} &  \begin{array}{l}h = h'\bullet x \points l\bullet l \points v \bullet y \points w \end{array} \\
\begin{array}{l}\top \end{array} & \begin{array}{l}\mbox{otherwise} \end{array}
\end{array}\right.\cr}
\]

\end{exa}

\begin{lemma}
\label{newmodellocal}The functions $new[x]$, $dispose[x]$, $mutate[x,v]$ and $lookup[x,y]$ are all local in the separation algebra $(H,\bullet,u_H)$ from example \ref{freesetmodel}.
\proof
Let $f = new[x]$ and assume $h'\#h$. We want to show $f(h'\bullet h) \sqsubseteq \{h'\} * f(h)$. Assume $h = h'' \bullet x \points v \bullet F$ for some $h''$, $x$, $l$, $v$ and $F$, because otherwise $f(h) = \top$ and we are done. So we have 
\[
\begin{array}{lll}
f(h'\bullet h) &=& \{h' \bullet h'' \bullet x \points l \bullet l \points w \bullet F\backslash \{l\}\mid w \in Val, l \in F\} \\
&=& \{h'\} * \{h'' \bullet x \points l \bullet l \points w \bullet F\backslash \{l\}\mid w \in Val, l \in F\}\\
&=& \{h'\} * f(h)
\end{array}
\]  
The other functions can be checked in a similar way. 
\qed
\end{lemma}

\subsection{Safety footprints for \emph{AD}}

We consider the footprint of the \emph{AD} command in the new model. In this model the sequential composition $new[x];dispose[x]$ gives the function
\[
AD(h)
= \left\{ \begin{array}{ll}
\begin{array}{l}\{h' \bullet x \points l \bullet F \mid l \in F\} \end{array} &  \begin{array}{l}h = h'\bullet x \points v \bullet F \end{array} \\
\begin{array}{l}\top \end{array} & \begin{array}{l}\mbox{otherwise} \end{array}
\end{array}\right. 
\]
The smallest safe states are given by the set $\{x \points v \bullet F \mid v \in Val, F \in \mathcal{P}(L)\}$. By lemma \ref{minstates}, these smallest safe states are footprints. However, unlike before, in this model these are the \emph{only} footprints of the $AD$ command. To see this, consider a larger state $h \bullet x \points v \bullet F$ for non-empty $h$. We have 
\[
\begin{array}{lll}
AD(h \bullet x \points v \bullet F) &=& \{h \bullet x \points l \bullet F \mid l \in F\}\\
&=& \{h\} * \{x \points l \bullet F \mid l \in F\}\\
&=& \{h\} * AD(x \points v \bullet F)
\end{array}
\]  
Since the local limit $L_{AD}(h \bullet x \points v \bullet F) \sqsubseteq \{h\} * AD(x \points v \bullet F)$ by definition, we have by proposition \ref{locallimitproposition} that $L_{AD}(h\bullet x \points v \bullet F) = AD(h\bullet x \points v \bullet F)$, and so $h\bullet x \points v \bullet F$ is not a footprint of $AD$. 

Thus the footprints of $AD$ in this model do not include any non-empty heaps. By corollary \ref{smallspeccorollary}, in this model the $AD$ command has a smallest complete specification in which the pre-condition only describes the empty heap. This specification is 
$$\{(\{x \points v \bullet F\}, \{x \points l \bullet F\}) \mid v \in Val, F \in \mathcal{P}(L), l \in F\}$$

Intuitively, it says that if initially the heap is empty, the variable $x$ is present in the stack, and we know which cells are free in the global heap, then after the execution, the heap will still be empty, exactly the same cells will still be free, and $x$ will point to one of those free cells. This completely describes the behaviour of the command for all larger states using the frame rule. For example, we get the complete specification on the larger state in which 42 is allocated: 
$$\{(\{42 \points w\} * \{x \points v \bullet F\}, \{42 \points w\}*\{x \points l \bullet F\}) \mid v,w \in Val, F \in \mathcal{P}(L), l \in F\}$$

In the pre-condition, the presence of location 42 in the heap means that 42 is not in the free set $F$ (by definition of $*$). Therefore, in the post-condition, $x$ cannot point to 42. 

Notice that in order to check that we have `regained'  safety footprints, we only needed to check that the footprint definition (definition \ref{def:footprint}) corresponds to the smallest safe states. The desired properties such as essentiality, sufficiency, and small specifications then follow by the results established in previous sections.

\subsection{Safety footprints for arbitrary programs}

Now that we have regained the safety footprints for \emph{AD} in the new model, we want to know if this is generally the case for \emph{any program}. We consider the abstract imperative programming language given in \cite{COY07}:
\[\begin{array}{rcl}
C & ::= &  c \mid \mathtt{skip} \mid  C;C \mid C+C \mid C^\star 
\end{array}
\]
where $c$ ranges over an arbitrary collection of primitive commands,
$+$ is nondeterministic choice, $;$ is sequential composition, and $(\cdot)^\star$
is Kleene-star (iterated $;$). As discussed in \cite{COY07}, conditionals and while loops can be encoded using $+$ and $(\cdot)^\star$ and assume statements. The denotational semantics of commands is given in Figure~\ref{fig:densemantics}.

\begin{figure}[t]
\begin{center}
\hrule
$$
\begin{array}{c}
\sem{c} \in LocFunc  \quad \quad 
\sem{\mathtt{skip}}(\sigma)= \{\sigma\} \\[1.5ex]
\sem{C_1;C_2} =  \sem {C_1};\sem{C_2} \quad \quad
\sem{C_1+C_2}  = \sem{C_1}  \sqcup \sem {C_2} \quad \quad
\sem{C^{\star}}   = \bigsqcup_n \sem{C^{\,n}}
\end{array}
$$

\end{center}
\hrule
\caption{Denotational semantics for the imperative programming language}
\label{fig:densemantics}
\end{figure}

Taking the primitive commands to be $new[x]$, $dispose[x]$, $mutate[x,v]$, and $lookup[x,y]$, our original aim was to show that, for every command $C$, the footprints of $\sem{C}$ in the new model are the smallest safe states. However, in attempting to do this, we identified a general condition on primitive commands under which the result holds for arbitrary separation algebras. 

Let $f$ be a local function on a separation algebra $\Sigma$. If, for $A \in \mathcal{P}(\Sigma)$, we define $f(A) = \displaystyle\bigsqcup_{\sigma \in A} f(\sigma)$, then the locality condition (definition \ref{def:localaction}) can be restated as
\[
\forall \sigma', \sigma \in \Sigma.\; f(\{\sigma'\}*\{\sigma\}) \sqsubseteq \{\sigma'\} * f(\{\sigma\})
\] 

The $\sqsubseteq$ ordering in this definition allows local functions to be more deterministic on larger states. This sensitivity of determinism to larger states is apparant in the \emph{AD} command in the standard model from example \ref{locsepexamples}.2. On the empty heap, the command produces an empty heap, and reassigns variable $x$ to \emph{any} value, while on the singleton cell 1, it disallows the possibility that $x = 1$ afterwards. In the new model, the $AD$ command does not have this sensitivity of determinism in the output states. In this case, the presence or absence of the cell 1 does not affect the outcomes of the $AD$ command, since the command can only assign $x$ to a value chosen from the free set, which does not change no matter what additional cells may be framed in. With this observation, we consider the general class of local functions in which this sensitivity of determinism is not present. 

\begin{defi}[Determinism Constancy]
\label{def:detconst}
Let $f$ be a local function and $\mathit{safe}(f)$ the set of states on which $f$ does not fault. $f$ has the determinism constancy property iff, for every $\sigma \in \mathit{safe}(f)$,
\[
\forall \sigma' \in \Sigma.\; f(\{\sigma'\}*\{\sigma\}) = \{\sigma'\} * f(\{\sigma\})
\] 
\end{defi}

Notice that the determinism constancy property by itself implies that the function is local, and it can therefore be thought of as a form of `strong locality'. 
Firstly, we find that local functions that have determinism constancy always have footprints given by the smallest safe states. 

\begin{lemma}
\label{detconstfootprint} If a local function $f$ has determinism constancy then its footprints are the smallest safe states.
\proof
Let $min(f)$ be the smallest safe states of $f$. These are footprints by lemma \ref{minstates}. For any larger state $\sigma'\bullet\sigma$ where $\sigma \in min(f)$, $\sigma' \in \Sigma$ and $\sigma$ is non-empty, we have 
\[
f(\sigma'\bullet \sigma) = f(\{\sigma'\}*\{\sigma\}) = \{\sigma'\} * f(\sigma) 
\]
Since $L_f(\sigma'\bullet\sigma) \sqsubseteq   \{\sigma'\} * f(\sigma)$, by proposition \ref{locallimitproposition} we have that $L_f(\sigma'\bullet\sigma) = f(\sigma'\bullet\sigma)$, and so $\sigma'\bullet\sigma$ is not a footprint of $f$. 
\qed
\end{lemma}

We now demonstrate that the determinism constancy property is preserved by all the constructs of our programming language. This implies that if all the primitive commands of the programming language have determinism constancy, then the footprints of every program are the smallest safe states.

\begin{thm}
\label{detconstprograms} If all the primitive commands of the programming language have determinism constancy, then the footprint of every program is given by the smallest safe states.
\end{thm}
\proof
Assuming all primitive commands have determinism constancy, we shall show by induction that every composite command has determinism constancy and the result follows by lemma \ref{detconstfootprint}. So for commands $C_1$ and $C_2$, let $f = \sem{C_1}$ and $g = \sem{C_2}$ and assume $f$ and $g$ have determinism constancy. For sequential composition we have, for $\sigma \in \mathit{safe}(f;g)$ and $\sigma' \in \Sigma$,  
\[\quad\eqalign{
&(f;g)(\{\sigma'\}*\{\sigma\})\cr
=\,\enspace&g(f(\{\sigma'\}*\{\sigma\}))\cr
=\,\enspace&g(\{\sigma'\}*f(\{\sigma\}))
\qquad\!\vbox to 7 pt{\baselineskip=12 pt\noindent%
  ($f$ has determinism constancy and
  \phantom{(}$\sigma\in\mathit{safe}(f)$ since 
  $\sigma\in\mathit{safe}(f;g)$)\vss}\cr
=\,\enspace&g(\bigsqcup_{\sigma_1 \in f(\sigma)} \{\sigma'\}*\{\sigma_1\})\cr
=\,\enspace&\bigsqcup_{\sigma_1 \in f(\sigma)} g(\{\sigma'\}*\{\sigma_1\})\cr
=\,\enspace&\bigsqcup_{\sigma_1 \in f(\sigma)} \{\sigma'\}* g(\sigma_1)
\qquad\vbox to 7 pt{\baselineskip=10 pt\noindent%
  ($g$ has determinism constancy and\\
  \phantom{(}$\sigma_1\in\mathit{safe}(g)$ since $\sigma\in\mathit{safe}(f;g)$
           and $\sigma_1\inf(\sigma)$)}\cr
=\,\enspace&\{\sigma'\}*\bigsqcup_{\sigma_1\in f(\sigma)}g(\sigma_1)%
  \qquad\mbox{(distributivity)}\cr
=\,\enspace&\{\sigma'\}* (f;g)(\sigma)\cr}
\]  

\noindent For non-deterministic choice, we have for $\sigma \in \mathit{safe}(f + g)$ and $\sigma' \in \Sigma$,  
\[\quad\eqalign{
&(f + g)(\{\sigma'\}*\{\sigma\})\cr
=\,\enspace&f(\{\sigma'\}*\{\sigma\}) \sqcup g(\{\sigma'\}*\{\sigma\})\cr
=\,\enspace&\{\sigma'\}*f(\{\sigma\}) \sqcup \{\sigma'\}*g(\{\sigma\})
\qquad\vbox to7 pt{\baselineskip=10 pt\noindent%
  ($f$ and $g$ have determinism constancy and \\
  \phantom{(}$\sigma \in \mathit{safe}(f)$ and $\sigma \in \mathit{safe}(g)$
  since $\sigma \in \mathit{safe}(f+g)$)}\cr\cr
=\,\enspace& \{\sigma'\}* (f(\{\sigma\}) \sqcup g(\{\sigma\}))\quad \mbox{(distributivity)}\cr
=\,\enspace& \{\sigma'\}* (f+g)(\{\sigma\})\cr}
\]  

\noindent For Kleene-star, we have for $\sigma \in \mathit{safe}(f^{\star})$ and $\sigma' \in \Sigma$,  
\[\quad\eqalign{
&(f^{\star})(\{\sigma'\}*\{\sigma\})\cr
=\,\enspace& \displaystyle \bigsqcup_{n} f^{n}(\{\sigma'\}*\{\sigma\})\cr
=\,\enspace& \displaystyle \bigsqcup_{n} \{\sigma'\}*f^{n}(\{\sigma\})
\qquad\vbox to7 pt{\baselineskip=10 pt\noindent%
  (determinism constancy preserved under sequential composition and \\
  \phantom{(}$\sigma \in \mathit{safe}(f^n)$)}\cr
=\,\enspace& \{\sigma'\}*\bigsqcup_{n} f^{n}(\{\sigma\})
  \qquad \mbox{\mathstrut(distributivity)}\cr
=\,\enspace& \{\sigma'\}* (f^{\star})(\{\sigma\})\rlap{\hbox to320 pt{\hfill}\qEd}}
\]\medskip

\noindent Now that we have shown the general result, it remains to check that
all the primitive commands in the new model of section \ref{altmodel}
do have determinism constancy.

\begin{proposition}
\label{rammodelsprop} Let $H_1$ be the stack and heap model of example \ref{locsepexamples}.2 and $H_2$ be the alternative model of section \ref{altmodel}. The commands $new[x]$, $mutate[x,v]$ and $lookup[x,y]$ all have determinism constancy in both models. The $dispose[x]$ command has determinism constancy in $H_2$ but not in $H_1$.  
\end{proposition}
\proof
We give the proofs for the new and dispose commands in the two models, and the cases for mutate and lookup can be checked in a similar way. For $dispose[x]$ in $H_1$, the following counterexample shows that it does not have determinism constancy. 
\[
\begin{array}{lll}
&& dispose[x](\{l\points v\}*\{x\points l\bullet l \points w\})\\
&=& dispose[x](\emptyset)\\
& = &\emptyset \\
&\sqsubset&  \{l\points v \bullet x\points l\}\\
& =& \{l\points v\}* dispose[x](x\points l\bullet l \points w)
\end{array}
\]
\noindent For $new[x]$ in $H_1$, any safe state is of the form $h \bullet x \points v$. For any $h' \in H_1$, we have 
$$\{h'\} * new[x](h \bullet x\points v) =  \{h'\} * \{ h \bullet x \points l \bullet l \points w \mid w \in Val, l \in L \backslash loc(h)\} \quad (\dagger)$$

If $h'\bullet h \bullet x \points v$ is undefined then $h'$ shares locations with $loc(h)$ or variables with $var(h) \cup \{x\}$. This means that the RHS in $\dagger$ is the empty set. We have $new[x](\{h'\}*\{h \bullet x\points v\}) = new[x](\emptyset) = \emptyset = \{h'\}* new[x](h \bullet x\points v)$. If $h'\bullet h \bullet x \points v$ is defined, then 
\[
\begin{array}{lll}
&& new[x](\{h'\}* \{h \bullet x\points v\})\\
&=& new[x](h' \bullet h \bullet x\points v)\\
&=& \{ h' \bullet h \bullet x \points l \bullet l \points w \mid w \in Val, l \in L \backslash loc(h'\bullet h)\}\\
&=& \{ h'\} * \{h \bullet x \points l \bullet l \points w \mid w \in Val, l \in L \backslash loc(h'\bullet h)\}\\
&=& \{ h'\} * \{h \bullet x \points l \bullet l \points w \mid w \in Val, l \in L \backslash loc(h)\}\\
&=& \{h'\} * new[x](h \bullet x\points v)  
\end{array}
\]
\noindent For $dispose[x]$ in $H_2$, any safe state is of the form $h \bullet x \points l\bullet l \points v \bullet F$. Let $h' \in H_2$.  We have 
$$ \{h'\} * dispose[x](h \bullet x \points l\bullet l \points v \bullet F) =  \{h'\} * \{ h \bullet x \points l \bullet F\cup\{l\}\} \quad (\dagger\mbox{\!}\dagger)$$

If $h'\bullet h \bullet x \points l\bullet l \points v \bullet F$ is undefined then either $h'$ contains a free set or it contains locations in $loc(h) \cup \{l\}$ or variables in $var(h) \cup \{x\}$. If $h'$ contains a free set or it contains locations in $loc(h)$ or variables in $var(h) \cup \{x\}$, then the RHS in $\dagger\mbox{\!}\dagger$ is the empty set. If $h'$ contains the location $l$ then also the RHS in $\dagger\mbox{\!}\dagger$ is the empty set since the free set $F \cup \{l\}$ also contains $l$. Thus in both cases the RHS in $\dagger\mbox{\!}\dagger$ is the empty set, and we have $dispose[x](\{h'\}*\{h \bullet x \points l\bullet l \points v \bullet F\}) = \emptyset = \{h'\} * dispose[x](h \bullet x \points l\bullet l \points v \bullet F)$.

If $h'\bullet h \bullet x \points l\bullet l \points v \bullet F$ is defined then we have 
\[
\begin{array}{lll}
&& dispose[x](\{h'\}*\{h \bullet x \points l\bullet l \points v \bullet F\})\\
&=& dispose[x](h' \bullet h \bullet x \points l\bullet l \points v \bullet F)\\
&=& \{h' \bullet h \bullet x \points l \bullet F\cup\{l\}\}\\
&=& \{h'\} * \{ h \bullet x \points l \bullet F\cup\{l\}\}\\
&=&  \{h'\} * dispose[x](h \bullet x \points l\bullet l \points v \bullet F) 
\end{array}
\]

\noindent For $new[x]$ in $H_2$, any safe state is of the form $h \bullet x \points v \bullet F$. Let $h' \in H_2$.  We have 
$$ \{h'\} * new[x](h \bullet x \points v \bullet F) =  \{h'\} * \{ h \bullet x \points l \bullet l \points w \bullet F\backslash\{l\} \mid w \in Val, l \in F\} \quad (\dagger\mbox{\!}\dagger\mbox{\!}\dagger)$$

If $h'\bullet h \bullet x \points v \bullet F$ is undefined then either $h'$ contains a free set or it contains locations in $loc(h)$ or variables in $var(h) \cup \{x\}$. In all these cases the RHS in $\dagger\mbox{\!}\dagger\mbox{\!}\dagger$ is the empty set, and so we have $new[x](\{h'\}*\{h \bullet x \points v \bullet F\}) = \emptyset = \{h'\} * new[x](h \bullet x \points v \bullet F)$.

If $h'\bullet h \bullet x \points v \bullet F$ is defined then we have 
\[
\begin{array}{lll}
&&new[x](\{h'\}*\{h \bullet x \points v \bullet F\})\\
&=& new[x](h' \bullet h \bullet x \points v \bullet F)\\
&=& \{h' \bullet h \bullet x \points l \bullet l \points w \bullet F\backslash\{l\} \mid w \in Val, l \in F\} \\
&=& \{h'\} * \{ h \bullet x \points l \bullet l \points w \bullet F\backslash\{l\} \mid w \in Val, l \in F\}\\
&=& \{h'\} * new[x](h \bullet x \points v \bullet F)
\end{array}
\]
\qed

Thus theorem \ref{detconstprograms} and proposition \ref{rammodelsprop} tell us that using the alternative model of example \ref{freesetmodel}, the footprint of every program is given by the smallest safe states, and hence we have regained safety footprints for all programs. In fact, the same is true for the original model of example \ref{locsepexamples}.2 if we do not include the dispose command as a primitive command, since all the other primitive commands have determinism constancy. This, for example, would be the case when modelling a garbage collected language \cite{Park06}.

\section{Conclusions}

We have developed a general theory of footprints in the abstract setting of local functions that act on separation algebras. Although central and intuitive concepts in local reasoning, the notion of footprints and small specifications had evaded a formal general treatment until now. The main obstacle was presented by the \emph{AD} problem, which demonstrated the inadequacy of the safety footprint notion in yielding complete specifications. In addressing this issue, we first investigated the notion of footprint which does not suffer from this inadequacy. Based on an analysis of the definition of locality, we introduced the definition of the footprint of a local function, and demonstrated
that, according to this definition, the footprints are the only essential elements necessary to obtain a complete
specification of the function. For well-founded resource models, we showed that the footprints are also sufficient, and we also presented results for non-well-founded models.

Having established the footprint definition, we then explored the conditions under which the safety footprint does correspond to the actual footprint. We introduced an alternative heap model in which safety footprints are regained for \emph{every} program, including \emph{AD}. We also presented a general condition on local functions in arbitrary models under which safety footprints are regained, and showed that if this condition is met  by all the primitive commands of the programming language, then safety footprints are regained for every program. The theory of footprints has proven very useful in exploring the situations in which safety footprints could be regained, as one only needs to check that the smallest safe states correspond to the footprint definition \ref{def:footprint}. This automatically gives the required properties such as essentiality and sufficiency, which, without the footprint definition and theorems, would need to be explicitly checked in the different cases.    

Finally, we comment on some related work. The discussion in this paper has been based on the static notion of footprints as {\em states} of the resource on which a program acts. A different notion of footprint has recently been described in \cite{HO08}, where footprints are viewed as {\em traces} of execution of a computation. O'Hearn has described how the {\em AD} problem is avoided in this more elaborate semantics, as the allocation of cells in an execution prevents the framing of those cells. Interestingly, however, the heap model from example \ref{freesetmodel} illustrates that it is not essential to move to this more elaborate setting and incorporate dynamic, execution-specific information into the footprint in order to resolve the {\em AD} problem. Instead, with the explicit representation of free cells in states, one can remain in an extensional semantics and have a purely static, resource-based (rather than execution-based) view of footprints.

\section*{Acknowledgement}
  The authors wish to thank Cristiano Calcagno, Peter O'Hearn and Hongseok Yang for detailed discussions on footprints. Raza
acknowledges support of an ORS award. Gardner acknowledges support of
a Microsoft Research Cambridge/Royal Academy of Engineering Senior
Research Fellowship.


\end{document}